\documentclass[aps,prd,twocolumn,superscriptaddress,nofootinbib,longbibliography,floatfix]{revtex4-2}
\usepackage{amsmath,amssymb,graphicx,bm}
\usepackage[colorlinks=true,linkcolor=blue,citecolor=blue,urlcolor=blue]{hyperref}

\newcommand{\kms}{\,\mathrm{km\,s^{-1}\,Mpc^{-1}}}
\newcommand{\LCDM}{$\Lambda$CDM}
\newcommand{\Mpbar}{\bar M_{\mathrm P}}
\newcommand{\DS}{\Delta S}
\makeatletter
\newcommand{\partheading}[2]{\par
 \if@nobreak\everypar{}\else\addpenalty\@secpenalty\addvspace{0.8cm \@plus 1ex \@minus .2ex}\fi
 {\interlinepenalty\@M\centering\normalfont\bfseries PART #1: #2\@@par}%
 \nobreak\vskip 0.5cm\relax\@afterindenttrue\@afterheading}
\makeatother
\AtBeginDocument{\raggedbottom}

\begin{document}

\title{Energy injection versus scaling symmetry as solutions to the Hubble tension after ACT DR6 and DESI DR2}

\author{Asantha Cooray}
\affiliation{Department of Physics and Astronomy, University of California, Irvine, CA 92697, USA}
\author{Alessandro Melchiorri}
\affiliation{Department of Physics, Sapienza University of Rome, Piazzale Aldo Moro 5, 00185 Rome, Italy}

\date{\today}

\begin{abstract}
We test two ways to raise the Hubble constant inferred from the cosmic microwave background, energy injection before recombination
and a scaling symmetry of the cosmological observables, using Planck, ACT DR6, CMB lensing, DESI DR2 baryon acoustic oscillations and
Pantheon+ supernovae, with and without the SH0ES calibration. Without SH0ES, minimal thermo-coupled early dark energy gives
$H_0=69.2\kms$ at its best fit; at $H_0=71\kms$ its effective $\chi^2$ lies $16.2$ above the \LCDM{} minimum, while
axion-like early dark energy lies $8.7$ below it. Planck polarization at $200\le\ell<400$ accounts for most of the penalty in
the thermo-coupled model. The scaling symmetry, realized with a mirror dark sector and correlated changes in the fine-structure
constant $\alpha$ and the electron mass, admits fits at $H_0\simeq72$--$75\kms$ after refitting the cosmological parameters. At the
scaling benchmark $\lambda=1.077$, the synchronous and exact-copy dark-atom models give a cosmic age of $12.75$\,Gyr. Relative
to \LCDM{}, their effective $\chi^2$ changes by $-1.7$ and $-3.3$ without SH0ES, and by $-31.4$ and $-33.0$ with
SH0ES. For a thermal mirror sector present during nucleosynthesis at this benchmark, the best-fit change in $\alpha$ is
$-2.43\%$ when only $\alpha$ varies and $-3.65\%$ at fixed hydrogen binding energy, for our baseline inputs. Both
shifts have the opposite sign to the recombination shift. Electron--positron annihilation can drive the latter through a scalar
field. Equivalence-principle tests then require a switch of the coupling, here a second field. In this two-field example the origin
of the damping remains open, as does the origin of the mirror baryon asymmetry if the sector is produced after nucleosynthesis.
The data thus do not exclude the scaling symmetry, but a complete model that realizes it is still missing.
\end{abstract}

\maketitle

%=====================================================================================================
\section{Introduction}\label{sec:intro}
%=====================================================================================================

The Cepheid-calibrated distance ladder gives $H_0=73.04\pm1.04\kms$~\cite{Riess2022}, and JWST has
confirmed the Cepheid photometry~\cite{Riess2024crowding,Riess2024jwst,Riess2025}. Calibrations with
the tip of the red giant branch and with J-region asymptotic giant branch stars give lower values,
$68$--$70\kms$~\cite{Freedman2025}. Within \LCDM{}, cosmic microwave background (CMB) data give
$H_0\approx67$--$68\kms$~\cite{Planck2018params,Louis2025}, and baryon acoustic oscillation (BAO) data
from DESI DR2, calibrated with BBN, give $68.5\pm0.6\kms$~\cite{DESI2025}. For reviews see
Refs.~\cite{Bernal2016,DiValentino2021,Schoneberg2022,Kamionkowski2023,CosmoVerse2025,Cai2026,Schoneberg2026a,Schoneberg2026b}.

The CMB fixes the angular acoustic scale $\theta_s=r_s/D_A$ to a relative precision of
$3\times10^{-4}$~\cite{Planck2018params}, and BAO data fix $hr_d$, with $r_d$ the sound horizon at the baryon
drag epoch, together with $\Omega_m$~\cite{KnoxMillea2020,Aylor2019}, so a higher $H_0$
needs a smaller sound horizon. Early dark energy (EDE) and extra radiation shrink it by raising the
expansion rate before recombination~\cite{Karwal2016,Poulin2019,Smith2020,Poulin2023}; a heavier
electron or a larger fine-structure constant does so by making recombination
earlier~\cite{Hart2018,Hart2020,Sekiguchi2021}. Both also change the damping tail, the early integrated
Sachs--Wolfe effect and the inferred $\Omega_mh^2$, which high-$\ell$ CMB and BAO data
constrain~\cite{Hill2020,Ivanov2020EDE,Calabrese2025,Cai2026}. Most such solutions therefore reach only
$H_0\approx69$--$70\kms$ without the SH0ES
calibration~\cite{Schoneberg2022,Schoneberg2026a,Schoneberg2026b,Cai2026}.

There is one known exception. If every gravitational free-fall rate $\sqrt{G\rho_i}$ and the
photon--electron scattering rate are rescaled by a common factor $\lambda$, dimensionless CMB
observables are almost unchanged~\cite{Zahn2003,CyrRacine2022}. The measured CMB temperature breaks
this symmetry, because it fixes the photon density~\cite{Fixsen1996,Ivanov2020T0}. Cyr-Racine, Ge and
Knox~\cite{CyrRacine2022} showed that a mirror dark sector can restore it, by supplying photon-like
radiation that is tightly coupled to dark baryons; they raised the Thomson rate by lowering the helium
fraction to $Y_p\simeq0.17$, which conflicts with observation. Zhang and Frieman~\cite{Zhang2023} let
the fine-structure constant differ at decoupling instead. Greene and Cyr-Racine showed that a joint
variation of $\alpha$ and $m_e$ scales the Thomson rate at fixed hydrogen binding
energy~\cite{Greene2023}, and built a model with atomic dark matter and varying
constants~\cite{Greene2024}, for which Planck 2018, pre-DESI BAO and SH0ES gave
$H_0=73.80\pm1.02\kms$. The most recent systematic comparison of proposed solutions notes that the
mirror world may be constrained by big-bang nucleosynthesis (BBN) and leaves it for future
work~\cite{Schoneberg2026b}.

In this paper we confront one model of each class with the same likelihoods: thermo-coupled early dark energy
(TCEDE)~\cite{Kamionkowski2025}, an energy-injection model in which a coupling to thermal relics supplies the potential, and the scaling symmetry. We find that
TCEDE, for which this is to our knowledge the first test against the full CMB, BAO and supernova likelihoods, does not raise
$H_0$ beyond about $70\kms$; axion-like EDE fitted in the same pipeline reproduces Ref.~\cite{Poulin2026}, so the likelihood
pipeline is not the cause. The scaling symmetry, which Ref.~\cite{Greene2024} found to be flat on earlier data, remains flat up
to $H_0\approx75\kms$ with ACT DR6 and DESI DR2. We then show that a mirror sector present at nucleosynthesis requires
$\alpha$ to be lower at that epoch, opposite in sign to its shift at recombination, and that a light scalar kicked at
$e^\pm$ annihilation generates the latter. We also quantify what the production of the mirror sector after BBN must deliver. Negative results apply only to the model
families that we searched.

Unless stated otherwise, we assume a spatially flat universe with a cosmological constant at late times, adiabatic initial
perturbations with a power-law spectrum, the measured CMB temperature, and a single massive neutrino (of $0.05$ or $0.06$\,eV
in the two pipelines of Sec.~\ref{sec:pipelines}). In every fit we vary the six standard parameters together with the
parameters of the model. The paper is organized as follows. Section~\ref{sec:data} describes the data, the pipelines and the
statistical conventions. Part~I presents the results: Sec.~\ref{sec:tcede} treats TCEDE and axion-like EDE,
Sec.~\ref{sec:symmetry} the scaling symmetry and Sec.~\ref{sec:bbn} nucleosynthesis. Part~II explores what a physical realization
of the symmetry would need; its calculations do not add up to a complete model, and Sec.~\ref{sec:untested} states what
was not computed. It treats the light scalar in Sec.~\ref{sec:krm} and the production of the mirror sector in
Sec.~\ref{sec:production}. We discuss the results in Sec.~\ref{sec:discussion} and conclude in Sec.~\ref{sec:conclusions}. The
Supplemental Material contains the technical details.

%=====================================================================================================
\section{Data and methods}\label{sec:data}
%=====================================================================================================

\subsection{Likelihoods}\label{sec:likelihoods}
We use the following likelihoods, through Cobaya~\cite{Torrado2021}. For the low-$\ell$ CMB we use Planck 2018
TT~\cite{Planck2018like} and EE from the SRoll2 maps~\cite{Pagano2020}. At high $\ell$ we use the ``P-ACT''
combination~\cite{Louis2025}, that is, the Planck PR3 \texttt{plik-lite} likelihood cut to $\ell<1000$ in TT and $\ell<600$ in
TE and EE, together with the ACT DR6 foreground-marginalized CMB-only likelihood; ``Planck high-$\ell$'' below always means
this cut, which excludes the Planck damping tail. We add ACT DR6 and Planck PR4 CMB
lensing~\cite{Madhavacheril2024,Qu2024,Carron2022}, DESI DR2 BAO~\cite{DESI2025} and the Pantheon+
supernovae~\cite{Scolnic2022,Brout2022}, the last either uncalibrated (``without SH0ES'') or with the Cepheid-host distances of
SH0ES~\cite{Riess2022} (``with SH0ES'').

The ACT calibration $A_{\rm act}$ has a Gaussian prior $\mathcal N(1,0.003)$ and the ACT polarization
efficiency $P_{\rm act}$ is free within $[0.9,1.1]$. The Planck calibration is set equal to
$A_{\rm act}$; this differs from the convention of Ref.~\cite{Louis2025}, which gives the Planck calibration its own
prior. It assigns a single prior to an amplitude offset that both data sets prefer, and the size of that
offset differs from fit to fit.
Section~\ref{sec:calib} repeats the main fits with an independent prior $\mathcal N(1,0.0025)$ on the
Planck calibration. For the fits refitted in both conventions $\DS$ then changes by $-0.52$ to $-0.12$, and no conclusion changes. SPT-3G data~\cite{Camphuis2026} are not used.

With SH0ES, the absolute magnitude $M_B$ of the supernovae is a free parameter, which we profile at every
evaluation. The calibration enters through the Pantheon+ and SH0ES likelihood of Cobaya, which replaces
the model distance of each supernova in a Cepheid host by the Cepheid distance of that host and uses the
full covariance matrix of the combined sample~\cite{Brout2022,Riess2022}. Reference~\cite{Poulin2026}
models the SH0ES calibration as a Gaussian prior on $M_B$ instead, so its gain with SH0ES is not directly
comparable with ours; where we quote it (Sec.~\ref{sec:ede}) we say so.

We minimize the score
\begin{equation}
S=\sum_i\left(-2\ln\mathcal L_i\right)+\left(\frac{A_{\rm act}-1}{0.003}\right)^2
\end{equation}
and quote $\DS\equiv S_{\rm model}-S_{\Lambda\rm CDM}$, with both scores computed in the same pipeline, at
the same accuracy and with the same supernova data. The two supernova branches contain different data,
so scores are never compared across them.

\subsection{Pipelines}\label{sec:pipelines}
We use two pipelines, both built on
CLASS~\cite{Lesgourgues2011,Blas2011}. A third, with a proxy for the dark plasma built on the interacting dark matter and dark radiation of stock
CLASS~\cite{CyrRacine2016,Archidiacono2019}, serves as a
cross-check of the scaling symmetry and is described in Sec.~S2 of the Supplemental Material~\cite{SM}.

In the TCEDE pipeline, a modified CLASS v3.3.4 integrates the background and the linear perturbations of
the coupled scalar--neutrino system in synchronous gauge, in the conventions of Ref.~\cite{Ma1995}
(Sec.~S9 of the Supplemental Material). TCEDE and its
\LCDM{} reference both use one massive neutrino of $0.05$\,eV (the bare mass in TCEDE),
$N_{\rm ur}=2.046$, and linear matter power for CMB lensing. Both use $\omega_\gamma=2.47\times10^{-5}$, which corresponds to $T_{\gamma0}=2.7247$\,K, $0.03\%$ below the measured value, and a temperature $T_{\nu0}=(4/11)^{1/3}T_{\gamma0}$ for the massive neutrino. Section~\ref{sec:tcede} tests the linear-power choice. For the
axion-like EDE control (Sec.~\ref{sec:ede}) the same code evolves an uncoupled scalar with the potential
of Eq.~\eqref{eq:ede} instead, with the same neutrino and lensing settings.

The dark-atom pipeline, which we use for the scaling symmetry, is built on \texttt{class\_adm}~\cite{Bansal2023}, that is,
on CLASS v3.1 with atomic dark matter and varying constants, with the changes listed in Sec.~S1 of the Supplemental Material.\footnote{One change corrects an error that affects any comparison with \LCDM{}.
As distributed, \texttt{class\_adm} computes visible hydrogen recombination in HyRec's simpler
\texttt{EMLA2s2p} mode whenever a dark sector is present and the visible constants are varied, while a
\LCDM{} run uses the default \texttt{SWIFT} mode. At $\lambda=1.001$, where the mirror sector is
physically negligible, this raises the score by $21$ at fixed parameters. We restricted the simpler mode
to the dark sector, after which it scores within $0.04$ of \LCDM{}.} Both sectors recombine
with HyRec~\cite{Lee2020}, and there is one massive neutrino of $0.06$\,eV. The \texttt{SWIFT} mode
corrects for radiative-transfer effects with a function that was calibrated for the standard constants.
At $\lambda=1.077$ it differs from HyRec's full radiative-transfer mode by up to $0.16\%$ in the
ionization fraction, against $7\times10^{-5}$ for \LCDM{}; applying the difference as a correction changes
$\DS$ by $+0.20$ and $+0.01$ for the two choices of dark constants at fixed parameters
(Sec.~S1 of the Supplemental Material). Refitted with the correction, the two fits give $\DS=-1.57$ and $-3.37$, against $-1.67$ and $-3.30$ without it. Non-linear power uses the
HMcode of CLASS v3.1~\cite{Mead2015,Mead2016}, which CLASS flags as unreliable when interacting dark
matter is present.
Replacing it by Halofit or by linear power changes $\DS$ at $\lambda=1.077$ by $0.0$ and $-0.4$ at fixed parameters. CLASS v3.1 computes the lensing potential only up to $k\approx1.6\,{\rm Mpc}^{-1}$, which with HMcode misses part of the lensing at high multipoles ($10\%$ of $C_L^{\phi\phi}$ at $L=3000$). Extending the range to $5.3\,{\rm Mpc}^{-1}$ changes $S$ at the \LCDM{} best fit by $+1.1$ but the differences $\DS$ by at most $0.2$, always in favor of the models (Sec.~S1 of the Supplemental Material), so we keep the default. For the physical realization of the symmetry the modulus field, its
perturbations and its decay radiation are evolved inside this code (Sec.~\ref{sec:krm_cmb}). Fits at fixed $\lambda$ are minimized with
Py-BOBYQA~\cite{Cartis2019}.

\subsection{Statistical conventions}\label{sec:stats}
All results are optimizer-based best fits and profiles; we quote no posteriors and no Bayesian
evidences. Optimizers are local, so a best fit means the best point found.
Every fit in the two pipelines is restarted from its best point with a smaller trust region until a restart gains less than
$0.1$. In the TCEDE pipeline the fits were then closed in rounds, in which a least-squares quadratic fitted to the evaluations near the lowest score proposed steps and placed evaluations along the directions that the evaluations did not yet reach, until a round lowered the score by less than $0.03$; the axion-like EDE fits, the TCEDE fit at $H_0=71\kms$ with the independent Planck calibration and the TCEDE best fit with HMcode were instead closed by repeated independent optimizer runs. The last independent optimizer run from each fit with a check of its own, and from one fit of each remaining family, lowered the higher-precision score by at most $0.08$ (Sec.~S1 of the Supplemental Material). Fits that move far along the direction in which $A_se^{-2\tau}$ and the calibration parameters compensate (those with an independent Planck calibration, those in another code and the bracketing refits of Sec.~\ref{sec:requirements}) minimize the calibration parameters inside every evaluation and add a line search after each restart (Sec.~S1 of the Supplemental Material). Every dark-atom fit quoted here was also tested with a quadratic through its evaluations and with one such restart from its best point: 14 of the $27$ fits fell by more than $0.05$ (the largest by $0.82$) and are replaced, and the \LCDM{} fit to the SH0ES data fell by $0.39$ (Sec.~S1 of the Supplemental Material). We treat differences below about $0.3$ as not meaningful.
This is a numerical criterion: the calibration convention alone moves $\DS$ by as much as $0.52$ (Sec.~\ref{sec:calib}).

The scores of the TCEDE pipeline are re-evaluated at higher precision, which moves them by at most $0.04$. In the
dark-atom pipeline, $\DS$ changes by at most $0.06$ when the precision settings are tightened twice
(Sec.~S1 of the Supplemental Material). At the same \LCDM{} point and settings the two codes agree to $0.03$ in $S$ with linear lensing power and to $0.01$ with HMcode once the lensing potential is computed to the same wavenumber. With the default range of CLASS v3.1 the HMcode score of the dark-atom code
differs by $-1.32$ (Sec.~S1 of the Supplemental Material). As a common control we fitted \LCDM{} with half an extra free-streaming species in both pipelines: $\DS=+24.8$ in the TCEDE pipeline (linear lensing power) and $+25.0$ in the dark-atom pipeline (HMcode). The two best-fit points of the TCEDE pipeline, evaluated in the dark-atom code with linear lensing power, give $+24.6$.

We count as extra parameters those varied in the fit, and not those of a microphysical completion, which we fix at benchmark
values in the second part of the paper. Values of the Akaike criterion are nominal: they count parameters and carry no Gaussian
significance here, because the TCEDE coupling is unidentified in the \LCDM{} limit, $\lambda\ge1$ is a boundary, and the two
choices of dark constants are alternative models, not points of one family. Some tests are single evaluations with the
cosmological parameters held fixed. They are diagnostics, carry no statistical interpretation, and are labeled ``fixed''.

Best-fit parameters are listed in Sec.~S5 of the Supplemental Material.

\begin{table}[t]
\caption{$\DS$ with the Planck calibration tied to the ACT calibration (the convention of this paper) and
with an independent prior. \LCDM{} is refitted in both pipelines. In the rows marked ``refitted'' the cosmological parameters are refitted with
the independent prior; in the rows marked ``fixed'' they stay at the best fit of the tied
convention and only the calibration parameters are re-profiled, so the entry is an upper bound. The other
fits are listed in Sec.~S1 of the Supplemental Material.}
\label{tab:calib}
\begin{ruledtabular}
\begin{tabular}{lccc}
Fit & tied & independent & cosmology\\
\colrule
\multicolumn{4}{l}{TCEDE pipeline (linear lensing power)}\\
TCEDE best fit & $-3.20$ & $-3.47$ & fixed\\
TCEDE, $H_0=71$ & $+16.18$ & $+15.66$ & refitted\\
Axion-like EDE best fit & $-8.67$ & $-8.95$ & fixed\\
\colrule
\multicolumn{4}{l}{Dark-atom pipeline (HMcode)}\\
$\lambda=1.077$, synchronous & $-1.67$ & $-1.79$ & refitted\\
$\lambda=1.077$, exact copy & $-3.30$ & $-3.52$ & refitted\\
\end{tabular}
\end{ruledtabular}
\end{table}

\subsection{Sensitivity to the calibration convention}\label{sec:calib}
Table~\ref{tab:calib} compares our convention, in which the Planck calibration is tied to the ACT one,
with the convention of Ref.~\cite{Louis2025}, in which the Planck calibration has its own prior
$\mathcal N(1,0.0025)$~\cite{Planck2018like} and the score gains the term $[(A_{\rm planck}-1)/0.0025]^2$. With the independent prior the \LCDM{} score rises by $0.53$ in the TCEDE pipeline and by $0.38$ in the dark-atom pipeline after refitting ($1.15$ and $0.95$ at fixed cosmology): \LCDM{} prefers a common calibration near $1.003$, for which the tied convention charges one prior and the independent convention two, and the refit moves part of that offset into the amplitude $A_se^{-2\tau}$.
For the fits that we refitted in both conventions, TCEDE at $H_0=71\kms$ and the two dark-atom fits at $\lambda=1.077$, $\DS$ changes by $-0.52$, $-0.12$ and $-0.22$. For the other fits we have upper bounds only, because their cosmology is not refitted, or is refitted along one direction only (Sec.~S1 of the Supplemental Material), while that of \LCDM{} is; they lie between $-0.28$ and $+0.44$ from the tied values. We keep the tied convention for the numbers quoted in the rest of the paper.

\subsection{Use of AI tools}\label{sec:ai}
We used the Claude/Fable 5.1 and 5.2 models of Anthropic to assist with code development, testing, the numerical runs and the
preparation of the manuscript. We specified the models and the likelihood comparisons, and we checked the reported scores through the
optimizer restarts, the higher-precision evaluations and the comparisons between the pipelines on common models that are described in
Sec.~\ref{sec:stats}.

\partheading{I}{RESULTS}

%=====================================================================================================
\section{Thermo-coupled early dark energy}\label{sec:tcede}
%=====================================================================================================

\subsection{Model}\label{sec:tcede_model}
Kamionkowski and Mathur~\cite{Kamionkowski2025} proposed that a scalar coupled to a thermal relic acquires
from that coupling the effective potential that an EDE episode needs, in place of a steep potential imposed by hand. The
field starts to move when the coupled species becomes non-relativistic; that this happens close to matter--radiation
equality is set by the initial value of the field and is not explained by the model. Their minimal realization couples a scalar $\phi$ with no bare
potential to one neutrino species through a field-dependent mass,
\begin{equation}
m_\nu(\phi)=m_{\nu0}\left(1+\varepsilon\,\frac{\phi^2}{\Mpbar^2}\right),
\label{eq:tcede_mass}
\end{equation}
where $\Mpbar=(8\pi G)^{-1/2}$ is the reduced Planck mass, used throughout this paper.
Reference~\cite{Kamionkowski2025} uses $M_{\rm P}=G^{-1/2}$, so its coupling is $8\pi\varepsilon$. The
model is related to mass-varying neutrinos~\cite{Fardon2004,Ichiki2008} and
neutrino-assisted EDE~\cite{Sakstein2020,CarrilloGonzalez2021,deSouza2023,CarrilloGonzalez2023}.
Reference~\cite{Kamionkowski2025} and a sterile-neutrino extension~\cite{Chowdhury2026} work at the
background level and leave the likelihood analysis to future work.

A homogeneous change in mass leaves the comoving momenta $q$ unchanged, so the coupled neutrinos keep
their relic Fermi--Dirac distribution with energy $E=\sqrt{q^2+a^2m_\nu^2(\phi)}$. With
$\beta\equiv d\ln m_\nu/d\phi$, the background equations are
\begin{align}
\ddot\phi+3H\dot\phi&=-\beta\,(\rho_\nu-3p_\nu),\label{eq:tcede_kg}\\
\dot\rho_\nu+3H(\rho_\nu+p_\nu)&=\beta\,\dot\phi\,(\rho_\nu-3p_\nu).\label{eq:tcede_cons}
\end{align}
Early on, the field sits at $\phi_i$ and the coupled neutrino is heavier than today. Once it becomes
non-relativistic, the trace $\rho_\nu-3p_\nu$ drives $\phi$ toward zero, where $m_\nu=m_{\nu0}$, and
neutrino rest energy is converted into scalar kinetic energy $\rho_K=\dot\phi^2/2$, which obeys
\begin{equation}
\dot\rho_K+6H\rho_K=-\beta\,\dot\phi\,(\rho_\nu-3p_\nu).
\label{eq:tcede_kin}
\end{equation}
The kinetic energy redshifts as $a^{-6}$ only once the source on the right is negligible. In our fits
that happens after recombination: at the best fit the source is still $10\%$ of the friction
term $6H\rho_K$ at $z=1100$, and the field first crosses zero at $z\approx560$
($7\%$ and $z\approx670$ at $H_0=71\kms$). We define the transient excess as
the scalar kinetic energy plus the energy of the coupled neutrino in excess of a neutrino of mass
$m_{\nu0}$, and quote it as a fraction $f$ of the total density, which includes the excess.
Reference~\cite{Kamionkowski2025} plots the same excess as a fraction $f_\Lambda$ of the density of the \LCDM{}
model without it, so that $f=f_\Lambda/(1+f_\Lambda)$.

We set $m_{\nu0}=0.05$\,eV and fit $\varepsilon$ and $\phi_i^2/\Mpbar^2$ together with the six \LCDM{}
parameters. Section~S9 of the Supplemental Material gives the perturbation equations, initial conditions and numerical
tests: at the benchmark of Ref.~\cite{Kamionkowski2025} the neutrino energy and momentum equations and the total energy
equation hold to $1.4\times10^{-7}$ or better and the total momentum equation to $1.8\times10^{-5}$ (95th percentiles), the uncoupled limits
reproduce \LCDM{} exactly, and doubling the momentum nodes, the multipoles and the background steps and tightening the
tolerance fivefold changes $S$ by at most $0.04$. The
axion-like control of Sec.~\ref{sec:ede} tests the likelihood pipeline but not the coupled perturbations, so we
checked their size where lensing is sourced: for $z\le5$ the density contrast of the coupled neutrino is
within $0.7\%$ ($1.4\%$ at $H_0=71\kms$) of that of an uncoupled $0.05$\,eV neutrino, the
coupling is strong only at $z\gtrsim100$, where less than $0.4\%$ of $C_L^{\phi\phi}$ is sourced, and the coupled
terms change $C_L^{\phi\phi}$ by less than $4\times10^{-5}$ ($9\times10^{-5}$) for $L\le1000$. For the benchmark of Ref.~\cite{Kamionkowski2025} (coupling $3\times10^3$ and
$\phi_i=4\times10^{27}$\,eV in the units of that work, $m_{\nu0}=0.05$\,eV, $H_0=67.4\kms$ and $\omega_b=0.0224$, to which we add
$\omega_{\rm cdm}=0.120$, $N_{\rm ur}=2.046$ and the photon and neutrino temperatures of
Sec.~\ref{sec:pipelines}) our code gives a peak of $f_\Lambda=17.81\%$, that is
$f=15.12\%$, at $z=5480$. The curve in Fig.~2(b) of that work, which we read from the vector
paths of the published figure, peaks at $f_\Lambda=17.95\%$
at $z=5890$. The amplitudes agree to $0.8\%$ in $f_\Lambda$ ($0.7\%$ in $f$), and our peak lies
$7\%$ lower in redshift in either convention. Raising $\varepsilon\phi_i^2$ by $6.5\%$ in our code reproduces the published
curve to an rms of $1\times10^{-4}$ in $f_\Lambda$. We could not identify the origin of the offset; the fit absorbs it,
because $\varepsilon$ and $\phi_i$ are free.

\subsection{Results}\label{sec:tcede_results}

\begin{table}[t]
\caption{TCEDE, axion-like EDE (Sec.~\ref{sec:ede}) and \LCDM{} best fits. $\DS$ is each model minus \LCDM{} with the
same supernova data.}
\label{tab:tcede_free}
\begin{ruledtabular}
\begin{tabular}{lcccc}
Supernovae & Model & $S$ & $H_0$ & $\DS$\\
\colrule
without SH0ES & \LCDM & 2235.11 & 68.3 & --\\
without SH0ES & TCEDE & 2231.92 & 69.2 & $-3.20$\\
without SH0ES & axion-like EDE & 2226.44 & 71.0 & $-8.67$\\
with SH0ES & \LCDM & 2313.71 & 68.7 & --\\
with SH0ES & TCEDE & 2299.43 & 69.9 & $-14.29$\\
with SH0ES & axion-like EDE & 2278.96 & 72.4 & $-34.75$\\
\end{tabular}
\end{ruledtabular}
\end{table}

\begin{table}[t]
\caption{Change in each likelihood, model minus \LCDM, at the best fits of Table~\ref{tab:tcede_free}.}
\label{tab:tcede_breakdown}
\begin{ruledtabular}
\begin{tabular}{lcccc}
 & \multicolumn{2}{c}{TCEDE} & \multicolumn{2}{c}{axion-like EDE}\\
Likelihood & no SH0ES & SH0ES & no SH0ES & SH0ES\\
\colrule
Planck low-$\ell$ TT & $-0.75$ & $-1.03$ & $-0.74$ & $-0.91$\\
SRoll2 low-$\ell$ EE & $-0.28$ & $-1.00$ & $-0.42$ & $-1.12$\\
Planck high-$\ell$ & $+2.86$ & $+6.18$ & $-3.77$ & $-3.29$\\
ACT DR6 & $-3.54$ & $-3.98$ & $-1.63$ & $-1.18$\\
DESI DR2 & $+0.28$ & $+0.58$ & $-0.42$ & $+0.29$\\
Supernovae & $+0.01$ & $-12.34$ & $+0.36$ & $-25.61$\\
CMB lensing & $-1.08$ & $-1.73$ & $-1.39$ & $-2.02$\\
$A_{\rm act}$ prior & $-0.70$ & $-0.96$ & $-0.67$ & $-0.92$\\
\colrule
Total & $-3.20$ & $-14.29$ & $-8.67$ & $-34.75$\\
\end{tabular}
\end{ruledtabular}
\end{table}

Without the SH0ES calibration, TCEDE improves the score by $\DS=-3.20$ for two extra
parameters (Table~\ref{tab:tcede_free}), which the Akaike criterion nominally counts in favor of \LCDM{}
($\Delta{\rm AIC}=+0.80$); Wilks' theorem does not strictly apply, because $\varepsilon$ is
unidentified at $\phi_i=0$. The best fit has $H_0=69.2\kms$, $n_s=0.982$, $\varepsilon\simeq200$ and
$\phi_i^2/\Mpbar^2=0.343$. The coupled neutrino starts with a mass of $3.48$\,eV, and the transient peaks
at an excess of $f=2.2\%$ at $z=5081$, far below the $15.2\%$ of the illustrative benchmark of
Ref.~\cite{Kamionkowski2025} in the same convention ($f_\Lambda=18.0\%$ as plotted there). The gain is spread over ACT DR6, lensing, the low-$\ell$ data and the calibration prior, and is partly offset by a penalty of $2.9$ in the Planck high-$\ell$ data (Table~\ref{tab:tcede_breakdown}); DESI and the supernovae are neutral. With SH0ES, $\DS=-14.29$, but $86\%$ of that comes from the
supernova term, and the best fit moves only to $H_0=69.9\kms$.

These fits use linear matter power for the lensing likelihood. Re-evaluating
both best fits with HMcode~\cite{Mead2021} at the same parameters lowers the \LCDM{} score by $3.76$
and the TCEDE score by $1.46$, mostly (about three quarters) in the lensing likelihood, so $\DS$ without SH0ES
shrinks from $-3.20$ to $-0.90$ ($-1.03$ with the 2015 version of
HMcode~\cite{Mead2015,Mead2016}). The supernova term is unaffected, so the same shift would
take $\DS$ with SH0ES from $-14.3$ to about $-12.0$.
Refitting both models with HMcode gives $\DS=-1.4$ without SH0ES, with TCEDE at $H_0=69.2\kms$. HMcode is not calibrated for a neutrino whose mass changes in time, so this is a
sensitivity test; it shows that the small gain without SH0ES is within the uncertainty of the
non-linear modeling. The costs at fixed $H_0$ quoted next are an order of magnitude larger, and they
survive the change of lensing prescription. Refitted with HMcode, TCEDE at $H_0=71\kms$ lies $19.8$ above the \LCDM{} best fit with HMcode, against $16.2$ with linear power; the lensing likelihood accounts for $+2.3$ of the difference of $+3.7$.

\begin{figure}[t]
\includegraphics[width=\columnwidth]{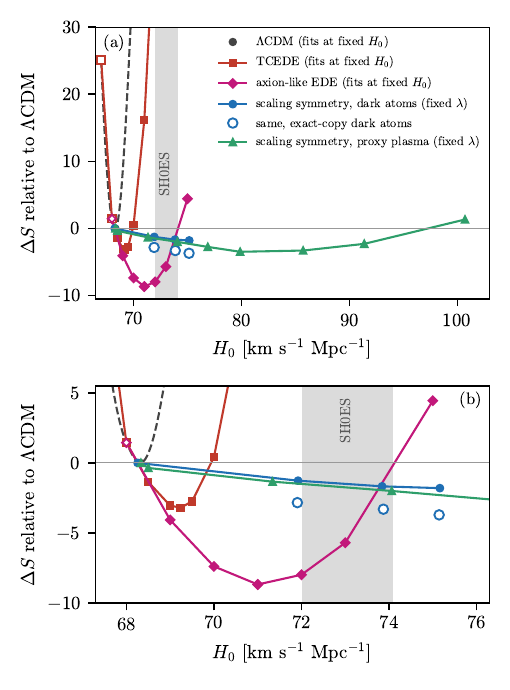}
\caption{Score relative to the \LCDM{} best fit of each pipeline, without the SH0ES calibration. TCEDE
(squares) and axion-like EDE (diamonds): best fit and fits at fixed $H_0$ in the same pipeline; open symbols:
field switched off (the model is \LCDM{}).
\LCDM: fits at fixed $H_0$, with the TCEDE fits at $67$ and $68\kms$ (scalar switched off), joined by a monotone interpolation. Scaling symmetry: fits at fixed $\lambda$, plotted at
their fitted $H_0$, with dark atoms (filled circles synchronous, open circles exact copy) and with the
proxy plasma of Sec.~S2 of the Supplemental Material (triangles; no supernovae). TCEDE, EDE and \LCDM{} use linear lensing
power and the dark-atom fits HMcode (Sec.~\ref{sec:pipelines}). The gray band is the SH0ES value. Panel (b) enlarges the range $H_0=67$--$76\kms$.}
\label{fig:profiles}
\end{figure}

At fixed $H_0$ and without SH0ES, the TCEDE score rises by $3.6$, $19.4$ and $50.2$ above its
own minimum at $H_0=70$, $71$ and $72\kms$ (Table~\ref{tab:tcede_profile}, Fig.~\ref{fig:profiles}).
Between the best fit and $70\kms$ it lies $0.4$ above its minimum at $H_0=69.5\kms$; below the best fit, $0.1$, $1.8$, $4.6$ and $28.3$ at $69$, $68.5$, $68$ and $67\kms$; at $68$ and $67\kms$ the scalar is switched off ($\phi_i=0$) and the profile is that of \LCDM{}.
The rise reaches $4$ at $H_0\approx70.0\kms$, against $\approx68.9\kms$ for \LCDM{}
(interpolating between the best fit and a fit at $H_0=70\kms$); with SH0ES the two values are $\approx70.6$ and $\approx69.2\kms$. TCEDE
therefore extends the upper end of the range of $H_0$ compatible with the data by about $1\kms$, but not to the local
value. Along the profile the transient grows from $2.2\%$ to $5.6\%$ ($4.5\%$ at $z=5790$ for
$H_0=71\kms$) and $n_s$ rises to $1.00$--$1.01$.
At $H_0=70$ and $71\kms$ the profile follows a single minimum; points with a smaller or a larger transient, where earlier fits had stopped, lie at least $1.0$ above it.

\begin{table}[t]
\caption{TCEDE fixed-$H_0$ profile: score relative to the TCEDE best fit, without and with the SH0ES
calibration. The first row is the TCEDE best fit, at $H_0=69.2\kms$ without and $69.9\kms$ with SH0ES; the
last three columns are for the fits without SH0ES. $f_{\rm peak}$ is the peak fractional energy excess.}
\label{tab:tcede_profile}
\begin{ruledtabular}
\begin{tabular}{lccccc}
$H_0$ & no SH0ES & SH0ES & $n_s$ & $\phi_i^2/\Mpbar^2$ & $f_{\rm peak}$\\
\colrule
best fit & 0 & 0 & 0.982 & 0.34 & 2.2\%\\
70 & $+3.6$ & $+0.2$ & 0.990 & 0.53 & 3.4\%\\
71 & $+19.4$ & $+9.4$ & 0.999 & 0.72 & 4.5\%\\
72 & $+50.2$ & $+35.6$ & 1.008 & 0.89 & 5.6\%\\
\end{tabular}
\end{ruledtabular}
\end{table}

\begin{figure*}[t]
\includegraphics[width=\textwidth]{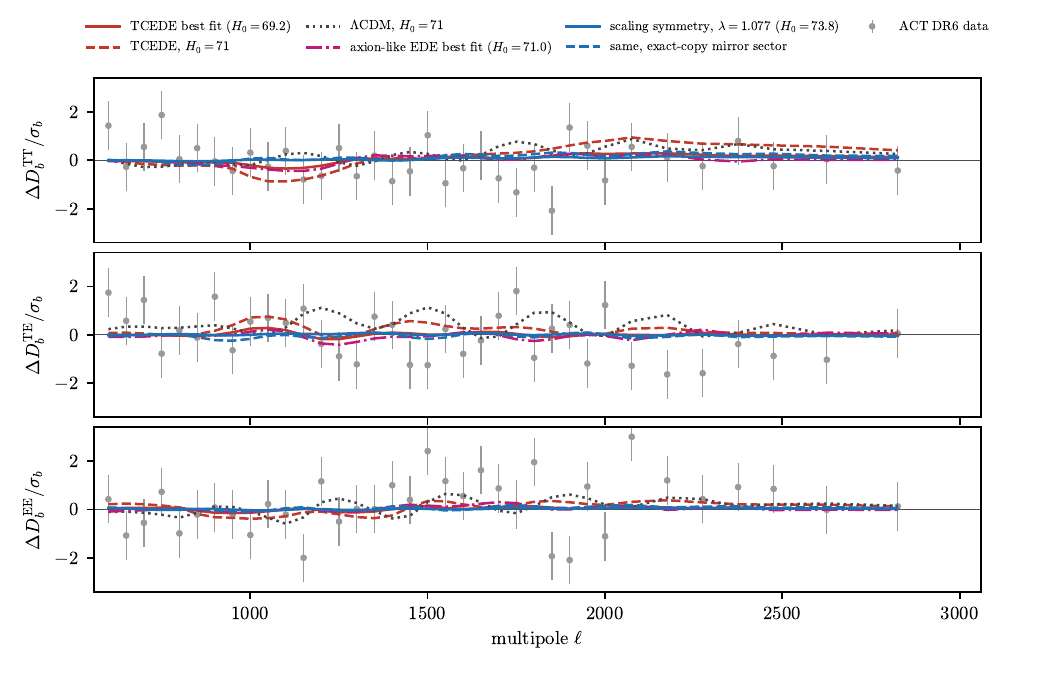}
\caption{Best-fit TT, TE and EE spectra (top to bottom) minus the \LCDM{} best fit of the same
pipeline, binned as the ACT DR6 CMB-only bandpowers and divided by their uncertainties $\sigma_b$.
Curves: TCEDE at its best fit and forced to $H_0=71\kms$; \LCDM{} forced to $H_0=71\kms$; axion-like EDE
at its best fit ($H_0=71.0\kms$); and the
scaling-symmetry model at $\lambda=1.077$ ($H_0=73.8\kms$) with synchronous and exact-copy dark
atoms. Points: the ACT DR6 bandpowers minus the \LCDM{} best fit. Each curve includes its fitted ACT
calibration and polarization efficiency. The \LCDM{} model forced to $H_0=71\kms$ is penalized mainly
by DESI and ACT DR6, and TCEDE forced to $71\kms$ by the Planck high-$\ell$ spectra, which are shown in
Fig.~\ref{fig:planckres} (Table~\ref{tab:h71}).}
\label{fig:residuals}
\end{figure*}

Table~\ref{tab:h71} compares \LCDM{} and TCEDE, both forced to
$H_0=71\kms$, with the \LCDM{} best fit, and Fig.~\ref{fig:residuals} shows the spectra. At that $H_0$
TCEDE is $82.2$ better than \LCDM{} but still $16.2$ worse than the \LCDM{} best fit. It removes
essentially the whole DESI, lensing and low-$\ell$ EE cost of the higher $H_0$ and most of the ACT DR6 and supernova cost, but
not the Planck high-$\ell$ cost, where it is worse than \LCDM{} at the same $H_0$ ($+16.7$ against
$+12.4$). With SH0ES, \LCDM{} at $71\kms$ gains only $9.3$ from the supernova term, because its low
$\Omega_m\simeq0.27$ costs about $10$ in the shape of the supernova Hubble diagram, whereas TCEDE keeps
$\Omega_m\simeq0.29$ and takes the full calibrator gain. These are profile differences, not
converted to significances.

\begin{figure*}[t]
\includegraphics[width=\textwidth]{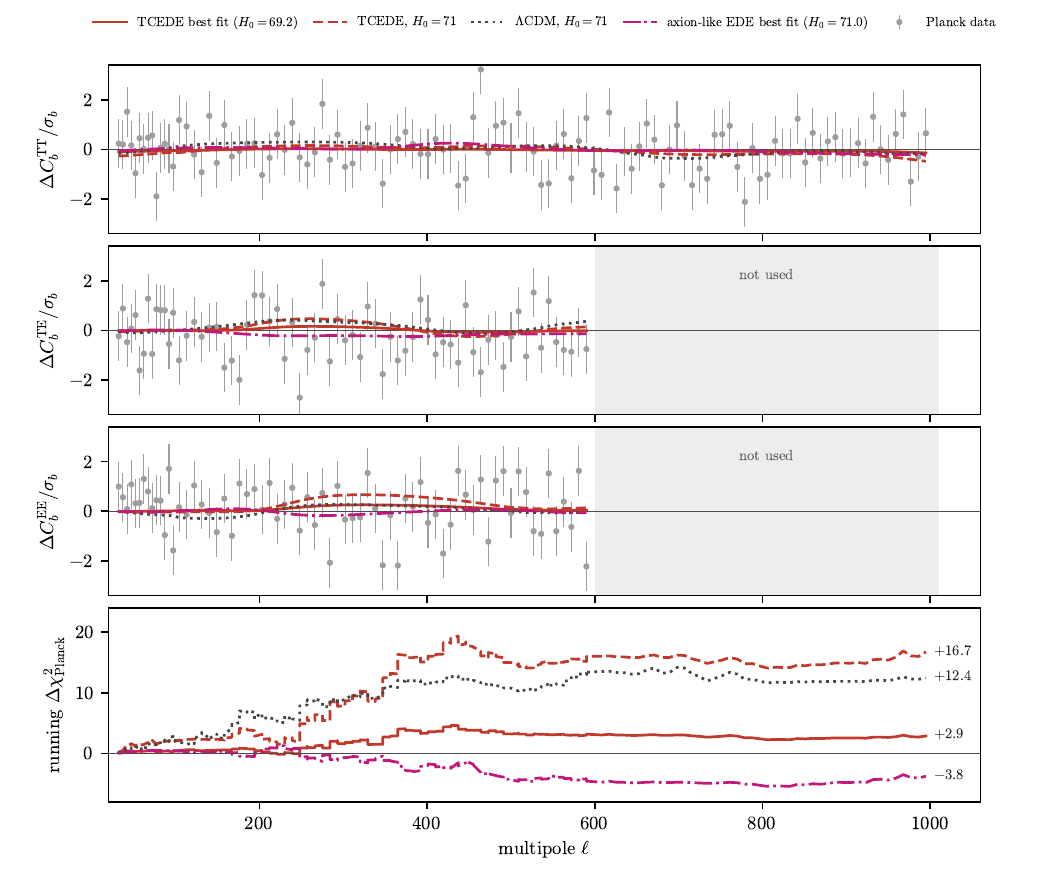}
\caption{The Planck PR3 \texttt{plik-lite} bandpowers used in the fits (TT at $\ell<1000$; TE and EE at
$\ell<600$) minus the \LCDM{} best fit of the TCEDE pipeline, divided by their uncertainties, with the
same TCEDE, \LCDM{} and axion-like EDE models as in Fig.~\ref{fig:residuals}. Each curve includes its
fitted calibration. The lowest panel shows, for each model, the running sum over multipoles of the
covariance-weighted contributions to its Planck $\chi^2$ relative to the \LCDM{} best fit; the sum of
the three spectra at the highest multipole is the entry of Table~\ref{tab:h71} or
Table~\ref{tab:tcede_breakdown}.}
\label{fig:planckres}
\end{figure*}

To identify the origin of the Planck penalty, we split the Planck high-$\ell$ term with its full covariance matrix $C$:
bin $i$ is assigned $r_i(C^{-1}r)_i$, where $r$ is the vector of residuals, so that the bins add up to
the $\chi^2$ and the correlations between bins and between spectra are kept. Of the $+16.7$ of TCEDE
at $H_0=71\kms$, EE carries $+10.3$, TE $+4.9$ and TT $+1.5$; evaluated alone, each with its own
block of the covariance, the three spectra change by $+12.0$, $+5.5$ and $+1.5$. The
multipoles $200\le\ell<400$ supply $12.2$ of the total: $8.0$ in EE, $4.8$
in TE and $-0.5$ in TT (Fig.~\ref{fig:planckres}). The penalty is therefore not a damping-tail effect, which the cuts at
$\ell<1000$ and $\ell<600$ exclude in any case, but comes from the Planck polarization spectra at
$\ell\approx200$--$400$. \LCDM{} forced to the same $H_0$ is penalized mostly in TE ($+9.1$, against
$+3.7$ in EE and $-0.4$ in TT), and the gain of axion-like EDE is also in TE
($-4.0$). At the TCEDE best fit the penalty of $2.9$ has the same pattern as at $71\kms$: $+1.6$
in EE, $+0.9$ in TE and $+0.3$ in TT.

\begin{table}[t]
\caption{Cost of $H_0=71\kms$ in each likelihood, relative to the \LCDM{} best fit with the same
supernova data. The totals are located to about $0.1$. Single terms are located less well, because a fit can slide along the direction in which the optical depth, the amplitude and the calibration compensate: closing the \LCDM{} fit at this $H_0$ lowered its total by $0.35$ but raised its low-$\ell$ EE term by $2.9$ ($3.1$ with SH0ES).}
\label{tab:h71}
\begin{ruledtabular}
\begin{tabular}{lcccc}
 & \multicolumn{2}{c}{without SH0ES} & \multicolumn{2}{c}{with SH0ES}\\
Likelihood & \LCDM & TCEDE & \LCDM & TCEDE\\
\colrule
DESI DR2 & $+37.4$ & $-1.8$ & $+39.5$ & $+0.2$\\
ACT DR6 & $+22.5$ & $+3.0$ & $+21.6$ & $+2.1$\\
Planck high-$\ell$ & $+12.4$ & $+16.7$ & $+11.8$ & $+16.1$\\
Supernovae & $+9.9$ & $+2.3$ & $-9.3$ & $-18.1$\\
CMB lensing & $+4.4$ & $-1.4$ & $+3.9$ & $-1.8$\\
SRoll2 low-$\ell$ EE & $+8.7$ & $-0.2$ & $+8.0$ & $-0.9$\\
$A_{\rm act}$ prior & $+4.1$ & $-0.7$ & $+3.9$ & $-0.9$\\
Planck low-$\ell$ TT & $-1.1$ & $-1.8$ & $-0.9$ & $-1.6$\\
\colrule
Total & $+98.4$ & $+16.2$ & $+78.5$ & $-4.9$\\
\end{tabular}
\end{ruledtabular}
\end{table}

\subsection{Axion-like EDE in the same pipeline}\label{sec:ede}
To separate the model from the pipeline, we fitted axion-like EDE~\cite{Poulin2019,Smith2020} to the same
data with the same code, settings and optimizer. A scalar with potential
\begin{equation}
V(\phi)=m^2f^2\left[1-\cos(\phi/f)\right]^3
\label{eq:ede}
\end{equation}
replaces the coupled one, and the neutrino is uncoupled. With a negligible scalar ($f_{\rm EDE}\sim10^{-8}$) the code reproduces the \LCDM{} scores of the TCEDE build to $2\times10^{-4}$ (at the \LCDM{} best fit, and at $H_0=68\kms$ at higher precision), but at standard precision its score at $68$ moves by up to $0.017$ ($0.13$ in single likelihoods) with the parameters of the scalar. At the
best fit of Ref.~\cite{Poulin2026}, with that work's neutrino mass and with HMcode, our pipeline
reproduces its $\chi^2$ to $0.5$ ($0.7$ per likelihood). We vary
$\log_{10}m$, $\log_{10}f$ and the initial angle $\theta_i=\phi_i/f$; the peak fraction $f_{\rm EDE}$ of
the total density and its redshift $z_c$ are derived.

Without SH0ES, EDE improves on \LCDM{} by $\DS=-8.67$ for three parameters ($\Delta{\rm
AIC}=-2.67$; Table~\ref{tab:tcede_free}) at $H_0=71.00\kms$, with $f_{\rm EDE}=0.094$,
$\log_{10}z_c=3.55$ and $\theta_i=2.70$. This is close to the best fit that
Ref.~\cite{Poulin2026} found with Planck at $\ell<1000$, ACT DR6, lensing, Pantheon+ and DESI DR2:
$H_0=70.98\kms$, $f_{\rm EDE}=0.090$, $\log_{10}z_c=3.549$ and $\theta_i=2.708$, with
$\Delta\chi^2=-7.0$. Most of the difference in the gain is in the lensing likelihood, $-1.39$
here against $+0.2$ there: with linear power, the higher matter density and amplitude of the EDE fit
are credited in lensing, as they were for TCEDE. Refitting EDE and \LCDM{} with HMcode, which reproduces the \LCDM{} lensing $\chi^2$ of Ref.~\cite{Poulin2026} ($19.6$ against $19.8$; $22.4$ with linear power), gives $\DS=-6.83$ at $H_0=70.92\kms$, with $+0.44$ in lensing, close to the $-7.0$ and $+0.2$ found there. Most of the EDE gain thus survives the change of lensing prescription, which reduces the gain of TCEDE from $3.2$ to $1.4$. With SH0ES, and with linear power
as for the $-14.29$ of TCEDE, $\DS=-34.75$ at $H_0=72.40\kms$. Reference~\cite{Poulin2026} finds $-35.4$ at $72.49\kms$ with
SH0ES entered as a Gaussian prior on $M_B$ and not through the Cepheid-host likelihood
(Sec.~\ref{sec:likelihoods}), so the two numbers are not like for like. A second start of the fit without SH0ES, from $\theta_i=1.5$, ends in a second minimum about $5$ above its best fit, at $\theta_i\approx1.3$.

The two models differ most in the Planck high-$\ell$ spectrum (Table~\ref{tab:tcede_breakdown}), which
EDE fits better than \LCDM{} by $3.8$ and TCEDE worse by $2.9$. At fixed $H_0=71$, $72$, $73$ and $75\kms$ EDE lies at $-8.7$, $-8.0$, $-5.7$ and $+4.4$ relative to the \LCDM{} best fit (Fig.~\ref{fig:profiles}), where TCEDE lies at $+16.2$ and $+47.0$ at $71$ and $72\kms$. The EDE score rises by $4$ above its minimum at $H_0\approx73.2\kms$, against $70.0\kms$ for TCEDE, and returns to the \LCDM{} best fit at $H_0\approx74.1$--$74.3\kms$, between the computed points at $73$ and $75\kms$. Below its best fit EDE lies at $-7.4$, $-4.1$ and $+1.4$ at $H_0=70$, $69$ and $68\kms$, where the field is switched off. The
pipeline therefore does not penalize energy injection as such; the shortfall of TCEDE is a property of
the model.

\subsection{Interpretation}\label{sec:tcede_interp}
On the current data the minimal model does not reach the values of $H_0$ needed to remove the tension.
The data prefer a small transient, the Planck polarization spectra at $\ell\approx200$--$400$ resist the
larger injection that a higher $H_0$ would need, and the SH0ES calibration does not change this. The fit also raises the baryon
density, from $\omega_b=0.0225$ in \LCDM{} to $0.0226$ at the TCEDE best fit and $0.0230$ at
$H_0=71\kms$, the direction that deuterium disfavors in early-time solutions~\cite{Giovanetti2026}.

Axion-like EDE, fitted the same way, does better (Sec.~\ref{sec:ede}); for other data combinations see
Refs.~\cite{Calabrese2025,Jhaveri2026,Schoneberg2026a,Schoneberg2026b}. The Planck penalty is a mismatch between polarization and temperature. With the TT spectrum matched to
$0.4\%$, the TCEDE spectra at $H_0=71\kms$ have $2.8\%$, $2.1\%$ and $1.5\%$
more EE power than the \LCDM{} best fit at $\ell=200$--$300$, $300$--$400$ and $400$--$500$
($1.1\%$, $0.8\%$ and $0.6\%$ at the TCEDE best fit), where the Planck data lie
slightly below \LCDM{}. Single evaluations at fixed acoustic angle show that the transient itself
produces the excess. Added to the \LCDM{} best fit with only the calibration refitted, it raises EE by
$8.6\%$, $6.3\%$ and $4.6\%$ in those bands and TT by at most $2.2\%$, and the
shifts of $\omega_{\rm cdm}$, $\omega_b$, $n_s$, $A_s$ and the calibration with which the fit restores TT remove
only $67$--$68\%$ of it, leaving the excess quoted above. An axion-like pulse of the same height and redshift raises EE relative to TT about a third as
much. The pulse of the EDE fit, about twice as high, raises it half as much as the TCEDE transient, and
the larger $\omega_{\rm cdm}$ of that fit cancels it. These evaluations suggest that the height and the
timing of the TCEDE transient are not what the data reject, and that what differs is its form. The
excess is broad and early: at the best fit it reaches half its peak at $z\approx1.4\times10^4$
and spans $1.9$ $e$-folds between its half-maximum points, against $0.8$ for the axion-like
fit; at $z=10^4$ it is at $73\%$ of its peak ($83\%$ at $H_0=71\kms$), where the
axion-like fit has $5\%$, and at recombination it retains $15\%$ of its peak
($11\%$), against $25\%$. And most of it, from two thirds at the peak to nine tenths when the modes at $\ell=500$ enter the horizon, is carried by the coupled neutrino, which is then semi-relativistic (rms speed $0.7$--$0.9\,c$). We
have not separated these two properties. The
conclusion therefore applies to this model, with its quadratic coupling, no bare potential and one
coupled species of $m_{\nu0}=0.05$\,eV, and not to energy injection in general. Other coupling
functions or potentials~\cite{CarrilloGonzalez2021,Baidya2026,SevillanoMunoz2026} and non-linear
clustering of the coupled neutrinos~\cite{Christiansen2026} were not considered.

%=====================================================================================================
\section{The scaling symmetry and a mirror dark sector}\label{sec:symmetry}
%=====================================================================================================

\subsection{The transformation and its implementation}\label{sec:sym_transform}
Under the scaling transformation with parameter $\lambda\ge1$, every energy density goes to
$\lambda^2\rho_i$, so every gravitational rate scales by $\lambda$; the Thomson rate $n_e\sigma_T$
also scales by $\lambda$, and the primordial amplitude goes to $A_s\lambda^{1-n_s}$. Dimensionless CMB
and BAO observables are then invariant up to atomic-physics corrections~\cite{Zahn2003,CyrRacine2022}.

The photon density is fixed by the measured CMB temperature~\cite{Fixsen1996,Ivanov2020T0}. Its missing
share, $(\lambda^2-1)\rho_\gamma$, must be supplied by dark photons that behave like visible ones: they
must be tightly coupled to dark baryons with the same baryon loading
$R'=3\rho_b'/4\rho_\gamma'=R$, and extra free-streaming radiation must keep the free-streaming fraction
unchanged~\cite{CyrRacine2022,Greene2024}. The Thomson rate is raised by varying the fine-structure
constant and the electron mass at fixed hydrogen binding energy
$B\propto\alpha^2m_e$~\cite{Greene2023,Greene2024}:
\begin{equation}
\alpha\to\lambda^{1/6}\alpha,\qquad m_e\to\lambda^{-1/3}m_e\quad\Rightarrow\quad\sigma_T\to\lambda\,\sigma_T .
\label{eq:rydberg}
\end{equation}
Starting from a \LCDM{} point we therefore apply $h\to\lambda h$,
$\omega_{\rm cdm}\to\lambda^2\omega_{\rm cdm}$, $N_{\rm ur}\to\lambda^2N_{\rm ur}$ and
$A_s\to\lambda^{1-n_s}A_s$, and add dark baryons with $\omega_b'=(\lambda^2-1)\,\omega_b$ and dark
photons with temperature ratio $\xi\equiv T'/T=(\lambda^2-1)^{1/4}$. The visible $\omega_b$ is
unchanged. For the massive neutrino we scale $(T_{\rm ncdm},m_{\rm ncdm})$ by $\lambda^{1/2}$, a
numerical device that scales its density by $\lambda^2$ at all times; it changes a laboratory mass,
which is harmless here because $\omega_\nu\simeq6\times10^{-4}$. The extra radiation is equivalent to
\begin{equation}
\Delta N_{\rm eff}=(\lambda^2-1)\left[N_{\rm eff}+\tfrac87\left(\tfrac{11}{4}\right)^{4/3}\right]\simeq7.45\,(\lambda^2-1),
\end{equation}
or $\Delta N_{\rm eff}\simeq1.2$ at $\lambda=1.077$, our benchmark.

The dark sector consists of dark hydrogen atoms, with two choices for their constants.
The \emph{synchronous} choice of Ref.~\cite{Greene2024}, $m_p'=\xi m_p$,
$m_e'=(\xi^4/\lambda)^{1/3}m_e$ and $\alpha'=(\lambda/\xi)^{1/6}\alpha$, makes dark recombination
coincide with the visible one, as the exact symmetry prescribes. The \emph{exact-copy} choice gives the
mirror particles the masses and coupling that the visible ones have at that epoch; dark recombination
then occurs earlier, at $z\approx1780$.

In this section the varied constants are prescribed: they return to their laboratory values in a step at
$z_t=50$, the default of the varying-constants implementation of CLASS, which Ref.~\cite{Greene2024} also uses. This is a placeholder for a physical history, and its details
do not matter: the history of a relaxing scalar field (Sec.~\ref{sec:krm}), prescribed in the same way, changes the score by
$0.02$ at identical parameters, and moving $z_t$ to $20$ or $300$ changes it by $0.1$
or less. At each fixed $\lambda$ we fit $\omega_b$, $\omega_{\rm cdm}$, $H_0$, $A_s$, $n_s$ and $\tau$,
with $A_{\rm act}$ and $P_{\rm act}$. The scaling parameter is therefore not the ratio of fitted Hubble
constants: the fitted $H_0$ ends $0.3$--$0.4\kms$ above $\lambda$ times the \LCDM{} value.

The background $H(z)/\lambda$ matches the base model to
$4\times10^{-5}$ at all redshifts. Recombination does not scale exactly: the Saha prefactor
$(m_eT)^{3/2}$ and the photoionization rates do not scale with $H$ when $m_e$ is lowered. Without a
refit, the image of the \LCDM{} best fit at $\lambda=1.077$ has $\theta_s$ shifted by $-0.17\%$ and
the redshift of last scattering, $z_\star$, by $+0.25\%$, spectra that differ from the base model by $1.0\%$ rms in TT and
$1.2\%$ in EE ($4.8\%$ and $3.2\%$ if the constants are not varied), and a score
$54$ above \LCDM{}; refitting removes this. Three further choices break the symmetry: the
visible helium fraction is held at the standard BBN value ($Y_{\rm He}=0.2454$, from the BBN table of CLASS), the mirror sector contains no helium, and the constants return at $z_t$.
The first two are tested below.

\subsection{Profile along the symmetry direction}\label{sec:sym_profile_text}

\begin{table}[t]
\caption{Fits along the symmetry direction at fixed $\lambda$, relative to \LCDM{} in the same code.
The SH0ES column evaluates each fit at its best point without SH0ES and compares it with the \LCDM{}
fit to the SH0ES data, so each entry is an upper bound on the value after a refit to the SH0ES data.
Fits to the SH0ES data at $\lambda=1.077$ end within $0.01$ of these entries.}
\label{tab:sym_profile}
\begin{ruledtabular}
\begin{tabular}{lcccc}
$\lambda$ & $H_0$ & $\DS$ & $\DS$ (SH0ES) & $n_s$\\
\colrule
\multicolumn{5}{l}{Synchronous dark constants}\\
1.000 & 68.3 & 0 & 0 & 0.974\\
1.050 & 71.9 & $-1.27$ & $-27.55$ & 0.972\\
1.077 & 73.8 & $-1.67$ & $-31.39$ & 0.971\\
1.095 & 75.2 & $-1.80$ & $-29.68$ & 0.971\\
\colrule
\multicolumn{5}{l}{Exact-copy dark constants}\\
1.050 & 71.9 & $-2.83$ & $-29.05$ & 0.970\\
1.077 & 73.9 & $-3.30$ & $-33.02$ & 0.969\\
1.095 & 75.1 & $-3.71$ & $-31.63$ & 0.968\\
\end{tabular}
\end{ruledtabular}
\end{table}

The symmetry direction is flat at the level of a few units of the score
(Table~\ref{tab:sym_profile} and Fig.~\ref{fig:profiles}). With synchronous dark atoms,
$\DS=-1.3$ to $-1.8$ over $H_0=72$--$75\kms$, and
$-27.5$ to $-31.4$ with SH0ES. The fitted parameters barely move: $n_s$ falls from $0.974$ to
$0.971$ and $\omega_b$ from $0.02254$ to $0.02250$. At $\lambda=1.077$ the changes relative to
\LCDM{} are $-0.64$ in Planck high-$\ell$, $-0.64$ in ACT DR6, $-1.19$ in DESI, $+0.85$ in the
low-$\ell$ data, $+0.26$ in Pantheon+, $-0.10$ in lensing and $-0.23$ in the calibration prior.
Unlike TCEDE, which loses $2.9$ in Planck high-$\ell$ at its best fit (Table~\ref{tab:tcede_breakdown}), the
model is not penalized by the high-$\ell$ CMB. The largest
term is DESI, which improves by $1.1$ to $1.5$ along the profile, because the refitted models
have $\Omega_m$ lower than \LCDM{} by $0.001$--$0.003$ and $hr_d$ higher by $0.2$--$0.4\%$.
With exact-copy dark atoms the fits are lower by a further $1.6$ to $1.9$:
$\DS=-3.3$ at $\lambda=1.077$, and $-33.0$ with SH0ES. The additional gain comes from the Planck
high-$\ell$ likelihood, which changes by $-2.7$ relative to \LCDM{}. Reference~\cite{Greene2024} found the same tendency on
earlier data when the epoch of mirror recombination was left free: a mild preference for mirror
recombination before the visible one ($\Delta\chi^2=-37.65$ against $-35.72$ with SH0ES).
At $\lambda=1.077$ we also fitted both models to the SH0ES data directly. The fits end within $0.01$ of the points fitted without SH0ES ($\DS=-31.4$ and $-33.0$, at $H_0=73.8$ and $73.9\kms$), so for these two rows the entries with SH0ES are fitted values.

A second implementation, with a proxy for the dark plasma in place of dark atoms and without
supernovae (Sec.~S2 of the Supplemental Material), gives the same picture over a wider range: $\DS=-2.0$ at
$H_0=74.1\kms$, a shallow minimum between the fits at $79.9$ and $85.7\kms$ ($-3.5$ and $-3.3$), and values within $3.84$ of that
minimum up to $H_0\approx98\kms$.

None of the differences without SH0ES amounts to a preference. The largest is
$3.7$ in the score for one scanned parameter, and the three sets of fits place their gains in different
likelihoods: DESI for synchronous dark atoms, Planck high-$\ell$ for the exact copy, and the high-$\ell$ CMB and
the calibration prior for the proxy
plasma. The difference between the two choices of dark constants is a difference between two dark
recombination histories, both computed with the simpler recombination model, which we did not test against a
full calculation (Sec.~S1 of the Supplemental Material). Above $75\kms$ the result rests on the proxy plasma alone, which omits the supernovae and
imposes a dark opacity that follows the visible one. The points are fits at fixed $\lambda$, not at
fixed $H_0$; each is an upper bound on the profile in $H_0$ at its fitted $H_0$. They establish that good
fits exist at these values of $H_0$. They do not give the depth of the global minimum or the shape of the
profile between the points.

On Planck 2018 and earlier BAO data,
Ref.~\cite{Greene2024} found $\Delta\chi^2=-1.08$ without supernovae, $-1.26$ with Pantheon+ and $-35.72$
with SH0ES, for synchronous dark atoms. ACT DR6 and DESI DR2 have not changed the picture: our
synchronous fits give $-1.3$ to $-1.8$ without SH0ES and $-27.5$ to $-31.4$ with it. The gain with
SH0ES is also similar to the
$-34.7$ of axion-like EDE in the TCEDE pipeline (Sec.~\ref{sec:ede}). Without SH0ES, axion-like EDE, with two more parameters, fits better than the symmetry at $72\kms$, where it lies at $-8.0$ against $-1.3$ and $-2.8$ for the synchronous and exact-copy fits at $71.9\kms$, and worse at $75\kms$, where it lies at $+4.4$ against $-1.8$ and $-3.7$ at $75.2$ and $75.1\kms$ (Fig.~\ref{fig:profiles}). We do not quote a crossing point. The two sets of points come from different pipelines and lensing prescriptions, the EDE points are fits at fixed $H_0$ and the symmetry points fits at fixed $\lambda$, and the change of lensing prescription alone moves the EDE gain by $1.8$ (Sec.~\ref{sec:ede}), which is comparable to the differences among the symmetry fits. The EDE profile rises because of the Planck high-$\ell$ and ACT DR6 terms: from the EDE best fit to $75\kms$ they grow by $7.3$ and $6.4$, and the total by $13.1$, almost all of it above $72\kms$. Along the symmetry profile, from \LCDM{} to the fits at $75.2$ and $75.1\kms$, the same two terms change by $-0.7$ and $-0.6$ for synchronous dark atoms and by $-2.9$ and $-1.0$ for the exact copy.

\subsection{Tests of the dark-sector ingredients}\label{sec:sym_demands}
We have tested two substitutes for the dark plasma at $\lambda=1.077$, with
the dark baryons counted as cold dark matter and the visible constants still shifted. If the dark photons
free-stream, the score at the best fit rises by $188$ (fixed, calibration parameters
re-profiled). Refitted in stock CLASS v3.3.4, this model lies $+13.7$ above \LCDM{} in the same code, at $H_0=73.8\kms$. A perfect-fluid dark radiation without drag does better but still fails: refitted, it lies $+9.4$ above \LCDM{}, at $H_0=74.2\kms$ (an earlier fit at $\lambda=1.095$, which had not converged, gave $+19.3$). Both refits meet the restart rule of Sec.~\ref{sec:stats}. These two substitutes therefore spoil the fits. They do not prove that
every dark sector without acoustic oscillations fails. The result quantifies, for our data, what
Ref.~\cite{Ge2023} showed at the level of the spectra, which change by several per cent when
free-streaming or fluid radiation replaces the mirror plasma.

Without the shift of the constants the image of \LCDM{} differs from it
three to five times more in the spectra (above); Ref.~\cite{Ge2023} showed at the level of the spectra that a mirror sector without a scaled photon scattering rate departs visibly from \LCDM{}.

With a mirror helium fraction equal to the visible one instead of zero, the score at
the best fit for $\lambda=1.077$ changes by $-0.1$ for synchronous constants and by
$0.0$ for the exact copy (fixed; a refit can only lower these values). The hydrogen-only
mirror sector therefore does not drive the results, and it is not what separates the two choices of
dark constants. Using the visible helium fraction that BBN predicts for the model of Sec.~\ref{sec:krm} instead of
the standard value changes the score by $+0.3$ at fixed parameters.

\subsection{Absolute time scales}\label{sec:clocks}
The symmetry leaves dimensionless observables unchanged, but it shortens every time interval by
$\lambda$, so observables that carry their own clock test it independently of the CMB and BAO data. SH0ES
is one. Stellar ages are another. This test is not new: Ref.~\cite{Ge2023} noted that lower limits on the
age at any redshift bound $\lambda$ from above, and applied the limit of Ref.~\cite{Vagnozzi2022}, which
includes the ages of high-redshift quasars. Ages are also an established test of any solution that
changes only the early Universe and reaches the SH0ES
value~\cite{Bernal2021,BoylanKolchin2021,Banik2026}. We add only the numbers for our fits
(Table~\ref{tab:ages}). The age falls from $13.77$\,Gyr for \LCDM{} to $12.75$\,Gyr at
$\lambda=1.077$, the same for the two choices of dark constants; the axion-like EDE best fit has
$13.3$\,Gyr. The most recent analysis of the oldest globular clusters gives a mean age of
$13.61\pm0.25\,({\rm stat})\pm0.23\,({\rm sys})$\,Gyr~\cite{Valcin2026}, where earlier ones gave
$13.3$\,Gyr~\cite{Valcin2020,Valcin2021}, and the oldest Galactic stars give
$13.80^{+0.20}_{-0.16}\,({\rm stat})\pm0.4\,({\rm sys})$\,Gyr~\cite{Banik2026}; see also
Ref.~\cite{Cimatti2023}. With the statistical and systematic errors added in quadrature
($0.34$\,Gyr for the clusters), the fit at $\lambda=1.077$ is $2.5\sigma$ younger than the
clusters if they formed at the big bang, and $3.1\sigma$ younger with the mean formation delay of
$0.20$\,Gyr found in Ref.~\cite{Valcin2026}; it is $2.4\sigma$ younger than the oldest
stars. As a one-sided test, requiring the Universe to be older than the clusters at $95\%$ confidence
gives $t_0>13.05$\,Gyr, or $H_0\lesssim72\kms$ along the symmetry direction, without a
formation delay, and $t_0>13.25$\,Gyr, or $H_0\lesssim71\kms$, with it. These
numbers depend on the treatment of the systematic error of the stellar models: with the two errors added
linearly the fit at $\lambda=1.077$ is $1.8\sigma$ low. Taken at face value, ages disfavor the
part of the symmetry direction that reaches the SH0ES value, which the CMB and BAO data leave open.

\begin{table}[t]
\caption{Ages $t_0$ of the fits along the symmetry direction (the two choices of dark constants agree to
$0.003$\,Gyr) and their distance below the mean age of the oldest globular clusters,
$13.61\pm0.34$\,Gyr~\cite{Valcin2026}, in units of that error, for clusters that formed at the big
bang and $0.20$\,Gyr after it.}
\label{tab:ages}
\begin{ruledtabular}
\begin{tabular}{lccccc}
 & \multicolumn{2}{c}{$H_0$} & & \multicolumn{2}{c}{below clusters ($\sigma$)}\\
$\lambda$ & synchr. & copy & $t_0$ (Gyr) & no delay & with delay\\
\colrule
1.050 & 71.9 & 71.9 & 13.09 & 1.5 & 2.1\\
1.077 & 73.8 & 73.9 & 12.75 & 2.5 & 3.1\\
1.095 & 75.2 & 75.1 & 12.53 & 3.2 & 3.8\\
\end{tabular}
\end{ruledtabular}
\end{table}

Intervals at high redshift shrink in the same way~\cite{BoylanKolchin2021}: the time between $z=20$ and
$z=7.6$, which is available for the growth of the first quasars~\cite{Nunes2020,Vagnozzi2022}, falls by
$7\%$, and unlike EDE~\cite{Shen2024} the symmetry does not form structure earlier to compensate.

%=====================================================================================================
\section{Nucleosynthesis: a mirror sector at BBN needs a lower $\alpha$}\label{sec:bbn}
%=====================================================================================================

An important question is whether the mirror sector, with $\Delta N_{\rm eff}\simeq1.2$, is already present
at BBN, since the answer determines which microphysics is allowed. We use PRyMordial~\cite{Burns2024} with its 12-reaction network and
PRIMAT rates~\cite{Pitrou2018}, repeat key cases with NACRE~II rates~\cite{Xu2013}, and extend it to
varying $\alpha$ and $m_e$ (Sec.~S3 of the Supplemental Material).

We compare the helium nucleon fraction $Y_p=4n_{\rm He}/n_b$ with $0.2458\pm0.0013$ from the Large Binocular Telescope (LBT) $Y_p$
project~\cite{Aver2026}, which quotes this quantity (the helium mass fraction used by the Boltzmann codes
is $0.54\%$ lower and is not compared with these data), and deuterium with $\mathrm{D/H}=(2.527\pm0.045)\times10^{-5}$, where the error
is the quadrature sum of the observational error, $0.030$~\cite{Cooke2018}, and a $1.3\%$ theory error.
The earlier value $Y_p=0.2453\pm0.0034$~\cite{Aver2021} leads to the same conclusions; a determination
from extremely metal-poor galaxies is lower, $0.2402\pm0.0040$~\cite{Yanagisawa2026}. Standard BBN at
the baryon density of our \LCDM{} fit, $\omega_b=0.02254$, gives $Y_p=0.2470$ and
$\mathrm{D/H}=2.416\times10^{-5}$ with PRIMAT, for which $\chi^2=6.9$
($6.1$ from deuterium and $0.8$ from helium), and
$\mathrm{D/H}=2.472\times10^{-5}$ ($\chi^2=2.2$) with NACRE~II. The two rate sets
differ by $2.3\%$ in D/H.

\subsection{A mirror sector present at BBN}\label{sec:bbn_mirror}
We evolve the mirror sector as a thermal copy ($\gamma'$, $e'^\pm$, three $\nu'$) at $\lambda=1.077$.
Its radiation is equivalent to $\Delta N_{\rm eff}=0.99$ at neutron--proton freeze-out and
$1.20$ by the end of nucleosynthesis. It gives $Y_p=0.2601$ and
$\mathrm{D/H}=2.80\times10^{-5}$, which is excluded ($\chi^2=159$). Keeping the
recombination-era constants unchanged back to BBN makes things worse ($\chi^2=297$).

A degeneracy exists: lowering $\alpha$ during BBN offsets the extra radiation in both $Y_p$ and D/H.
The shifts below are finite fractional changes of the constants, held fixed through BBN. Varying
$\alpha$ alone, the best fit is $\delta\alpha/\alpha=(-2.43\pm0.19)\%$ ($\chi^2=0.4$; the error is the half-width for $\Delta\chi^2=1$), with
$Y_p=0.2454$ and $\mathrm{D/H}=2.553\times10^{-5}$. At exactly fixed binding energy,
$(1+\delta\alpha/\alpha)^2(1+\delta m_e/m_e)=1$, it is $\delta\alpha/\alpha=(-3.65\pm0.31)\%$ with
$\delta m_e/m_e=+7.72\%$ ($\chi^2=0.0$), with $Y_p=0.2459$ and
$\mathrm{D/H}=2.523\times10^{-5}$. Table~\ref{tab:bbnshift} shows how the two solutions depend on
the nuclear rates, the helium measurement, the baryon density and the electromagnetic part of the
neutron--proton mass difference, $Q_{\rm QED}$ (Sec.~S3 of the Supplemental Material). With the earlier helium value the two baseline solutions are $-2.60\pm0.35\%$ ($\chi^2=0.1$) and $-3.66\pm0.51\%$ ($\chi^2=0.0$), with all pulls below $0.3$ in magnitude. The sign is the same in all $40$ combinations that we tried,
including two treatments of the theory error on D/H ($1.3\%$ and the full difference between the rate
sets, $2.3\%$). The size is $2.4$--$3.1\%$ for $\alpha$ alone and $3.5$--$4.2\%$ at
fixed binding energy; with $Q_{\rm QED}=-0.76$\,MeV the fixed-binding solution needs $4.5$--$5.7\%$
and fits worse ($\chi^2=1.5$--$5.0$). The direction of this
degeneracy was already noted in Ref.~\cite{Ichikawa2004}, and a value of $\alpha$ lower by about $2\%$ was shown to fit
the EMPRESS helium measurement~\cite{Matsumoto2022} without a change of the radiation density~\cite{Seto2023}. A thermal mirror sector
present at BBN would therefore need $\alpha$ to be $2.4$--$3.7\%$ \emph{lower} at BBN, for our
baseline inputs, and $1.2\%$ \emph{higher} at recombination.

\begin{table*}[t]
\caption{Shift of $\alpha$ during BBN that compensates a thermal mirror sector at $\lambda=1.077$, for
$\alpha$ varied alone and for $\alpha$ and $m_e$ varied at exactly fixed binding energy. Columns: best
shift, $\chi^2$ and the pulls of $Y_p$ and D/H (model minus data, in units of the error) with the LBT
helium value and a $1.3\%$ theory error on D/H; and the best shift with the earlier helium value,
$0.2453\pm0.0034$. The first row is the baseline (PRIMAT rates, $\omega_b=0.02250$,
$Q_{\rm QED}=-1.00$\,MeV); each other row changes one input. The change of $\omega_b$, $0.5\%$, is of
the size of the uncertainty of current CMB determinations~\cite{Louis2025}.}
\label{tab:bbnshift}
\begin{ruledtabular}
\begin{tabular}{lcccccccc}
 & \multicolumn{4}{c}{$\alpha$ alone} & \multicolumn{4}{c}{fixed binding energy}\\
Input & $\delta\alpha/\alpha$ (\%) & $\chi^2$ & pulls $Y_p$, D/H & earlier $Y_p$ & $\delta\alpha/\alpha$ (\%) & $\chi^2$ & pulls $Y_p$, D/H & earlier $Y_p$\\
\colrule
PRIMAT & $-2.43$ & $0.4$ & $-0.3$, $+0.6$ & $-2.60$ & $-3.65$ & $0.0$ & $+0.1$, $-0.1$ & $-3.66$\\
NACRE~II & $-2.53$ & $3.4$ & $-0.8$, $+1.7$ & $-2.94$ & $-3.83$ & $1.0$ & $-0.5$, $+0.9$ & $-4.19$\\
$\omega_b-0.00011$ & $-2.46$ & $1.1$ & $-0.5$, $+1.0$ & $-2.72$ & $-3.71$ & $0.1$ & $-0.2$, $+0.3$ & $-3.85$\\
$\omega_b+0.00011$ & $-2.40$ & $0.0$ & $-0.1$, $+0.2$ & $-2.48$ & $-3.59$ & $0.3$ & $+0.3$, $-0.5$ & $-3.47$\\
$Q_{\rm QED}=-0.76$\,MeV & $-3.09$ & $0.1$ & $+0.1$, $-0.2$ & $-3.05$ & $-5.28$ & $5.0$ & $+1.5$, $-1.7$ & $-4.46$\\
\end{tabular}
\end{ruledtabular}
\end{table*}

The conflict between the mirror sector and the light-element abundances is
known~\cite{CyrRacine2022,Zhang2023} (Ref.~\cite{Schoneberg2026b} notes it as a possible constraint), and the original proposals already suggest or assume
that the sector is reheated after BBN~\cite{CyrRacine2022,Zhang2023,Greene2023}; Ref.~\cite{Greene2023}
left the effect of the varied constants on BBN for future work. BBN limits on shifted
constants~\cite{Coc2007,Dent2007,Seto2023me,Garramone2026} and on radiation that appears between BBN
and recombination~\cite{Loverde2026} are also known, as is the fact that the helium yield is not invariant
under the scaling of the constants~\cite{Baryakhtar2024}. Here we add the size and the sign of the shift
that a thermal mirror sector at BBN would need; this does not exclude such a sector, but specifies the
shift of the constants that it would require.

\subsection{A light scalar as the source of the shift}\label{sec:bbn_scalar}
A light scalar with a linear coupling to $m_e$ is sourced by the electron trace, $\rho_e-3p_e$, which
is large while $T\sim m_e$, so it receives a kick as the pairs annihilate. Kicks at mass thresholds are
well known in scalar--tensor gravity~\cite{Damour1993,Damour1994,Catena2004,Coc2006} and for chameleon
fields~\cite{Brax2004,Erickcek2013}, and a thermal bath displaces a weakly coupled scalar quite
generally~\cite{Batell2023}; a coupling quadratic in the field instead gives the scalar a thermal
mass~\cite{Sibiryakov2020,Bouley2023,Baryakhtar2025}.

For a field that starts at rest the kick has a fixed sign: the field moves so as to lower the electron
rest energy, so $m_e$ decreases, and for a coupling that preserves the binding energy $\alpha$
increases. The timing is also fixed: $2\%$ of the kick is complete at $T=0.7$\,MeV,
$49\%$ at $0.1$\,MeV and $74\%$ at $0.05$\,MeV, so it happens during BBN and pushes
$\alpha$ in the wrong direction for the degeneracy above. One such field could give $\alpha$ lower by
$3\%$ at neutron--proton freeze-out and higher by $1.2\%$ at recombination only with a coupling
$\kappa\simeq0.19$ [Eq.~\eqref{eq:krm}], about twice that of Sec.~\ref{sec:krm}, and an initial displacement of
$-0.16\,\Mpbar$ tuned against it. Even then $\alpha$ would sweep through
$4.2\%$ during nucleosynthesis instead of staying low, and the equivalence-principle
violation discussed in Sec.~\ref{sec:ep} would be $3.3$ times larger. A single field of this kind
therefore does not supply the history that a mirror sector at BBN needs, and in what follows we take the
sector to be populated \emph{after} BBN, as suggested in Refs.~\cite{CyrRacine2022,Zhang2023}. This is a
choice within our construction, not a result. The calculations of this section leave open a history of
the constants that changes sign between BBN and recombination, a sector that is only partly populated at
BBN, and mirror baryons that are present early while most of the mirror radiation is supplied later.
We examine in Sec.~\ref{sec:production} what late population requires.

\partheading{II}{A PHYSICAL REALIZATION OF THE SYMMETRY}

In Part~I the scaling symmetry was treated phenomenologically. A physical realization needs an origin for
the shift of the constants, a mechanism that makes the shift disappear by today, and a way to create the
mirror sector after BBN. In this Part we study one line of attack on each. The
calculations do not add up to a complete model.

%=====================================================================================================
\section{A kicked Rydberg modulus}\label{sec:krm}
%=====================================================================================================

\subsection{The kick as the origin of the shift}\label{sec:krm_kick}
The kick that disfavors a scalar-induced shift at BBN (Sec.~\ref{sec:bbn}) has the sign that the
recombination shift needs, so we use it to \emph{generate} that shift. Consider a light modulus
$\varphi$ with
\begin{gather}
\mathcal L\supset-\tfrac12(\partial\varphi)^2-\tfrac12m^2\varphi^2,\nonumber\\
\alpha=\alpha_0\,e^{\kappa\varphi/\Mpbar},\qquad m_e=m_{e0}\,e^{-2\kappa\varphi/\Mpbar}.
\label{eq:krm}
\end{gather}
This direction preserves the Rydberg energy $\alpha^2m_e$ exactly, so hydrogen binding is unchanged, as
Eq.~\eqref{eq:rydberg} requires; all our calculations use these exponential forms. The field equation
also has a friction term $\Gamma\dot\varphi$. A decay into light dark fermions with Yukawa coupling $y$
would give $\Gamma=y^2m/8\pi$, but that width does not apply while $|\varphi|$ is large, because the
fermion mass $y\varphi$ then exceeds $m/2$. We therefore treat $\Gamma$ throughout as a phenomenological
rate at which the field hands energy and momentum to radiation, not as a computed decay width;
Sec.~S1 of the Supplemental Material gives the covariant form of the transfer. The
benchmark is $\Gamma=0.2\,m$; Secs.~\ref{sec:krm_cmb} and~\ref{sec:ep} show how the results depend on it. We assume that the field starts at $\varphi=0$, its present-day minimum, as it would with a
Hubble-induced mass during inflation. A light spectator would instead random-walk, and an initial
displacement would add a shift present at BBN.

In terms of $u=\varphi/\Mpbar$ and $x=\ln a$, during radiation domination
\begin{equation}
u''+\left(3+\frac{d\ln H}{dx}\right)u'=6\kappa\,\frac{\rho_e-3p_e}{\rho_{\rm tot}} ,
\label{eq:kick}
\end{equation}
where the electron trace is evaluated with the shifted mass. The kick takes place during BBN
(Sec.~\ref{sec:bbn}) and leaves a persistent displacement $u_\infty=c\,\kappa$, so
\begin{equation}
\ln\lambda=6\,c\,\kappa^2 .
\label{eq:lambda_kappa}
\end{equation}
In linear response $c=1.094$. The kick itself is not new: the same displacement is known for chameleon
fields, about $0.56\,\beta\,\Mpbar$ for masses that scale as $e^{\beta\varphi/\Mpbar}$~\cite{Erickcek2013},
which is $c\simeq1.1$ for $|\beta|=2\kappa$. The back-reaction of the shifted electron mass on the source raises $c$
to $1.102$, $1.106$ and $1.109$ for $\lambda=1.050$, $1.077$ and $1.095$, which correspond to
$\kappa=0.086$, $0.106$ and $0.117$ and to shifts $\delta\alpha/\alpha=0.82\%$, $1.24\%$
and $1.52\%$. We neglect the dependence of the plasma pressure on $\alpha$. At order $\alpha$ each
relativistic charged fermion lowers the pressure by $5\pi\alpha T^4/72$, which pushes the field the other
way for as long as such particles are present. With $e^\pm$ as the only charged species this lowers $c$
by about $0.12\%$ per $e$-fold of expansion, so the size of the effect depends on the temperature
at which the field is released: about $0.4\%$ for $10$\,MeV and $0.7\%$ for $100$\,MeV, and
more for a field that is free at higher temperatures, where muons, pions and quarks contribute. A change
of $c$ is absorbed by $\kappa$, which scales as $c^{-1/2}$ at fixed $\lambda$.

Equation~\eqref{eq:lambda_kappa} fixes the shift of the constants, and so the value of $\lambda$ that the
mirror sector must match, for one field that starts at rest at its minimum. It does not by itself fix
$H_0$. The mirror radiation has to be produced with the abundance that corresponds to the same $\lambda$:
Sec.~\ref{sec:requirements} varies it at fixed constants and finds $H_0$ between $72.1$ and $74.8\kms$.
The remaining parameters are fitted, and with two fields the relation between coupling and shift is a
different one (Sec.~\ref{sec:ep}).

After the kick $\varphi$ is frozen while $m\ll H$. For the benchmark mass, $m=6\times10^{-30}$\,eV, the
ratio $m/H$ is $0.17$ at $z=1100$ and reaches unity at $z\approx350$. The field
holds the shift through recombination ($0.5\%$ of it has relaxed by $z=800$), then relaxes:
the amplitude of $\delta\alpha/\alpha$ is $4\times10^{-4}$ at $z=50$ and $3\times10^{-6}$ at $z=20$
(Fig.~\ref{fig:alpha}; Fig.~S1 of the Supplemental Material is a timeline of the scenario).

\begin{figure}[t]
\includegraphics[width=\columnwidth]{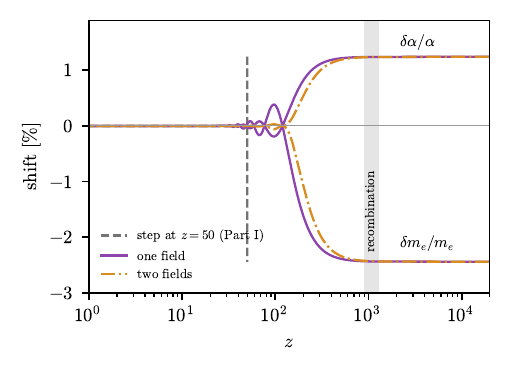}
\caption{Fractional shifts of $\alpha$ and $m_e$ in the one-field benchmark ($\lambda=1.077$,
$m=6\times10^{-30}$\,eV, $\Gamma=0.2\,m$) and in the two-field benchmark of Sec.~\ref{sec:ep}, as
evolved in the Boltzmann code, compared with the step at $z=50$ of Part I. The shaded band marks
recombination.}
\label{fig:alpha}
\end{figure}

\subsection{BBN with the kick}\label{sec:bbnkick}
With no mirror sector at BBN, we follow the kick through nucleosynthesis with the time-dependent
$\alpha(T)$ and $m_e(T)$ of the field solution (Sec.~S3 of the Supplemental Material), at the baryon density of the
corresponding CMB fit. At the benchmark the kick raises D/H by $2.3\%$ and $Y_p$ by
$0.0010$. With PRIMAT rates $\chi^2=4.0$, against $6.1$ for standard BBN at
the same baryon density; with NACRE~II rates, which start from a higher D/H, $\chi^2=2.7$
against $1.8$ (Fig.~S2 of the Supplemental Material). Helium contributes most of the benchmark's $\chi^2$: the
predicted $Y_p=0.2480$ lies $1.7\sigma$ above the LBT value. The two rate compilations
differ by as much as the effect of the kick, so BBN neither favors nor excludes the benchmark.

\subsection{CMB with the modulus in the Boltzmann code}\label{sec:krm_cmb}
The fits of Part I prescribe $\alpha(z)$ and ignore the energy of the field. Here the field is part of
the calculation: we evolve $\varphi$ with its damping, its linear perturbations in
synchronous gauge, and the radiation produced by the damping, which we treat as free-streaming and
source with the perturbed energy and momentum that the field loses (Sec.~S1 of the Supplemental Material). The
constants follow the evolved field through Eq.~\eqref{eq:krm}, and the helium fraction follows from
BBN with the kick.

The kick fixes the displacement, so the energy of the field is not a free
parameter. The field holds $\tfrac12m^2\varphi^2$ with $\varphi=0.117\,\Mpbar$ until it thaws, and
reaches $0.84\%$ of the total density at $z\approx170$; the fraction, unlike the redshift, does not
depend on its mass. It then passes
its energy to the decay radiation, which today has $0.5$ times the density of the CMB photons.
At the parameters of the Part~I fit this has two effects. It shortens the distance to last scattering,
which shifts $\theta_s$ by $+0.055\%$, $1.8$ times the precision of the measurement.
And, because neither the field nor its decay radiation clusters on the scales of interest, it slows the
growth of structure: $\sigma_8$ falls by $0.9\%$ and the lensing power by up to $3\%$.
Together they raise the score by $10$.

\begin{table}[t]
\caption{Fits at $\lambda=1.077$ with the modulus evolved in the Boltzmann code, for the benchmark
damping unless stated. $\DS$ is relative to \LCDM{}; the SH0ES column is formed as in
Table~\ref{tab:sym_profile}, so its entries are upper bounds. ``Prescribed'' is $\DS$ without SH0ES with the same dark constants and the
history of $\alpha$ and $m_e$ prescribed as in Part~I: the one-field history for synchronous constants
(its best fit lies within $0.01$ of that of the step in Table~\ref{tab:sym_profile}) and the step for the exact copy. The
two-field model is described in Sec.~\ref{sec:ep}. The last row is the exact copy with the fraction of the mirror radiation that
streams freely set to $F=0.478$, the value for decays into quark jets, instead of $0.409$ (Sec.~\ref{sec:requirements}).}
\label{tab:krm_fits}
\begin{ruledtabular}
\begin{tabular}{lcccc}
Model & $H_0$ & $\DS$ & $\DS$ (SH0ES) & prescribed\\
\colrule
one field, $\Gamma=m/8\pi$ & 73.5 & $+4.2$ & $-25.5$ & $-1.7$\\
one field & 73.6 & $+3.1$ & $-26.6$ & $-1.7$\\
two fields & 73.7 & $+1.4$ & $-28.3$ & $-1.7$\\
two fields, exact copy & 73.8 & $-0.7$ & $-30.4$ & $-3.3$\\
\quad with $F=0.478$ & 73.8 & $-0.5$ & $-30.2$ & $-2.8$\\
\end{tabular}
\end{ruledtabular}
\end{table}

How long the energy stays depends on $\Gamma$. With $\Gamma=m/8\pi$
($y=1$) the field oscillates as matter for several $e$-folds before it decays; the decay radiation today
has $1.1$ times the photon density, and the same evaluation raises the score by
$17$. With $\Gamma=0.5\,m$ the field is close to overdamped and keeps its potential energy
longer; the rise is $12$. The benchmark $\Gamma=0.2\,m$ is close to the minimum.

Refitting recovers part of the cost (Table~\ref{tab:krm_fits}). With the benchmark
damping the one-field fit has $\DS=+3.1$ without SH0ES and $-26.6$ with it, at
$H_0=73.6\kms$, against $-1.7$ and $-31.4$ with prescribed constants; with
$\Gamma=m/8\pi$ the values are $+4.2$ and $-25.5$. Relative to the prescribed history the largest changes are $+2.2$ in ACT DR6, $+1.2$ in DESI and $+0.8$ in lensing, while the fitted $H_0$ falls by $0.1\kms$ and $\Omega_m$ rises by $0.3\%$. Without SH0ES the model is therefore worse than \LCDM{}; with SH0ES it remains $27$ below it, because $H_0$ barely moves. These are the scores of the
benchmark at $\lambda=1.077$, not minima of the model. As $\kappa\to0$ the kick, the mirror sector
matched to it and the energy of the field all vanish, so the family contains \LCDM{}, and its best member
is at least as good as \LCDM{}.

\subsection{Late-time bounds}\label{sec:krm_late}
For the one-field benchmark the envelope of the residual oscillation of $\alpha$ is $2\times10^{-13}$ at
$z=7.1$, where quasar absorption systems at $z=5.5$--$7.1$ give a weighted mean of $(-2.2\pm7.3)\times10^{-5}$~\cite{Wilczynska2020}, and
$5\times10^{-23}$ at $z=4$. Below $z\approx2$ the oscillation has decayed and the residual is the static offset
sourced by today's electron density. It lies four or more orders of magnitude below the bounds from
quasar spectra at $z\simeq1$ ($10^{-6}$ in $\alpha$~\cite{Murphy2022}), from methanol absorption
($4\times10^{-7}$ in $m_p/m_e$~\cite{Kanekar2015}), from Oklo ($10^{-7}$~\cite{Damour1996}) and from atomic
clocks ($10^{-18}\,{\rm yr}^{-1}$~\cite{Lange2021}). All of these numbers depend on the damping term: with $\Gamma=m/8\pi$ the envelope at $z=7.1$ is
$1\times10^{-6}$, still below the quasar bound.
Planck and BAO data together limit shifts of $\alpha$ and $m_e$ at recombination to the per-cent
level~\cite{Hart2020}; the shifts used here, $+1.24\%$ and $-2.44\%$, are allowed only along the
degenerate direction that the mirror sector opens.

\subsection{Equivalence principle and a second field}\label{sec:ep}
The field of Eq.~\eqref{eq:krm} has a Compton wavelength of about $1$\,Mpc, so on
laboratory and planetary scales it mediates a long-range force between bodies, and MICROSCOPE constrains that force. In the dilaton
normalization of Damour and Donoghue~\cite{Damour2010a,Damour2010b}, Eq.~\eqref{eq:krm} gives couplings
$d_e=\sqrt2\kappa\simeq0.15$ and $d_{m_e}=-2d_e$, so a body $A$ couples through the combination
$Q_A\equiv Q_e-2Q_{m_e}$ of its dilaton charges, and
$\eta({\rm Ti,Pt})=d_e^2\,(Q_{\rm Ti}-Q_{\rm Pt})\,Q_\oplus$. For the alloys of the test masses and
for the Earth we find $Q_{\rm Ti}=1.74\times10^{-3}$, $Q_{\rm Pt}=3.74\times10^{-3}$ and
$Q_\oplus=1.41\times10^{-3}$, hence $\eta({\rm Ti,Pt})\simeq-2.8\times10^{-6}\,d_e^2$, or
$-6\times10^{-8}$ for the benchmark, against the
measured $(-1.5\pm2.7)\times10^{-15}$~\cite{Touboul2022}, with the statistical and systematic errors added in quadrature. The one-field model is therefore excluded by seven orders of magnitude, a margin that no plausible error in the
charges changes: with our formula, keeping $|\eta|$
below twice the measurement uncertainty requires
$d_e\lesssim4.4\times10^{-5}$, in line with
published limits on linear couplings of light scalars~\cite{Berge2018,Baryakhtar2025}. The axio-dilaton
realization of a varying electron mass~\cite{Smith2025axio} faces the same problem.

A second field can switch the coupling off; this is one way to do it, and we have not surveyed others. Let the Standard Model
couple only to the product $\varphi\sigma$, so that Eq.~\eqref{eq:krm} holds with
$\kappa\varphi/\Mpbar$ replaced by $\kappa(\varphi/\Mpbar)(\sigma/v_\sigma)$. If $\sigma$ sits at
$v_\sigma$ during the kick and later relaxes to zero, the matter coupling of $\varphi$ vanishes today.
The coupling of $\sigma$ is proportional to the value of $\varphi$ today, so both fields must relax.

The switch field is not a spectator during the kick: the electron trace sources both fields, nothing holds
$\sigma$ at $v_\sigma$, and for small $v_\sigma$ the two fields amplify each other
(Sec.~S7 of the Supplemental Material gives the equations and the kick for $v_\sigma/\Mpbar=0.02$--$0.3$). At
$v_\sigma=0.02\,\Mpbar$ the switch field grows by a factor $3.3$, the required coupling,
$\kappa=0.053$, is half that of the one-field model, and the shift scales as $\kappa^{3.9}$,
against $\kappa^2$ for one field, so Eq.~\eqref{eq:lambda_kappa} does not hold in the two-field model.

As a two-field benchmark we take $v_\sigma=0.02\,\Mpbar$, equal masses
$m_\sigma=m_\varphi=6\times10^{-30}$\,eV and $\Gamma=0.2\,m$ for both fields, start both from their
post-kick values, and evolve them in the Boltzmann code as in Sec.~\ref{sec:krm_cmb}. The shift of the
constants at recombination is that of the one-field model by construction (Fig.~\ref{fig:alpha}). The
two fields reach $0.32\%$ and $0.26\%$ of the total density at $z\approx170$, and at
the parameters of the Part~I fit the score rises by $6$; with $v_\sigma=0.05$ and $0.03\,\Mpbar$
the rise is $14$ and $8$, so the benchmark is one point of a family in which the
switch scale, the coupling and the energy of the fields are tied together. The kick is later than with one
field, and BBN gives $Y_p=0.2474$ and $\mathrm{D/H}=2.466\times10^{-5}$ with PRIMAT
($\chi^2=3.4$). Today both fields have decayed to below $10^{-20}$ of $v_\sigma$ and $\Mpbar$, so
their linear couplings to matter are negligible; this assumes that near the Earth the fields take their
cosmological values (Sec.~S7 of the Supplemental Material). Refitting gives $\DS=+1.4$ without SH0ES and $-28.3$ with it, at $H_0=73.7\kms$ (Table~\ref{tab:krm_fits}); with exact-copy dark constants the values are $-0.7$ and $-30.4$, and $-0.5$ and $-30.2$ for the quark-jet composition $F=0.478$.
The damping matters for the equivalence principle. If $\sigma$ were undamped, $\varphi$ would stay coupled at $d_e\approx5.5\times10^{-5}$, giving $\eta\approx-8.48\times10^{-15}$; if $\varphi$ were undamped, $\eta\approx-1.04\times10^{-14}$. These lie $2.6\sigma$ and $3.3\sigma$ from the measured central value, so both must be damped for the benchmark to lie within $2\sigma$.

\subsection{Naturalness and prior work}\label{sec:krm_prior}
Two aspects of the model are not natural. Electron loops correct the mass by
$\delta m^2\sim(2\kappa m_e/\Mpbar)^2m_e^2/16\pi^2\approx3\times10^{-36}\,{\rm eV}^2$, against
$m^2\approx4\times10^{-59}\,{\rm eV}^2$, a tuning of $23$ orders of magnitude that is generic to
ultralight scalars with Standard Model couplings~\cite{Batell2023,Cyncynates2024,Baryakhtar2025} (see Refs.~\cite{Uzan2011,Uzan2025} for reviews of varying constants). And the fields need strong
damping, $\Gamma\approx0.2\,m$, whose origin we do not specify.

Scalar-field realizations of a varying $m_e$ or $\alpha$ at recombination exist: ultralight and
hyperlight scalars~\cite{Zhang2023,Baryakhtar2024,Baryakhtar2025,Ghosh2026}, a
symmetron~\cite{Solomon2022}, dark-sector couplings~\cite{Hoshiya2023} and the
axio-dilaton~\cite{Smith2025axio}, reviewed in Ref.~\cite{Schoneberg2025}. In these the shift comes
from an initial displacement of the field or from its environment. The ingredients of the mechanism used
here are also known. Kicks at mass thresholds, including the size of the kick at $e^\pm$ annihilation,
were computed for scalar--tensor gravity and for chameleons (Sec.~\ref{sec:bbn}). Thermal displacement of
a scalar that starts at its minimum, in proportion to its coupling, has been used to set the abundance of
scalar dark matter~\cite{Batell2023,Cyncynates2024,Alachkar2024}, and a scalar driven by the electron
density changes $m_e$ in the theory of Ref.~\cite{Barrow2005}. A kick from neutrinos that become
non-relativistic is the trigger of neutrino-assisted EDE~\cite{Sakstein2020}. Specific to this work are the
use of the annihilation kick to generate the recombination-era shift along the Rydberg-preserving
direction, the resulting automatic sign, and the relation~\eqref{eq:lambda_kappa} between the coupling
and the shift, which fixes the $\lambda$ that the mirror sector must match. We did not find this use in a search of the literature, which covered the titles and abstracts
of the papers that cite Refs.~\cite{Greene2024,Baryakhtar2024} and the full text of
Refs.~\cite{Zhang2023,Baryakhtar2024,Baryakhtar2025,Ghosh2026,Smith2025axio,Schoneberg2025}; we do not
claim that no such work exists.

%=====================================================================================================
\section{Producing the mirror sector after BBN}\label{sec:production}
%=====================================================================================================

Following the choice made in Sec.~\ref{sec:bbn}, we take the mirror sector to be populated after BBN. What
the symmetry requires is the following.
The Universe after BBN must look as if a fraction $1-\lambda^{-2}$ of a denser photon--baryon--electron
plasma, all species in proportion, had been moved into a decoupled copy: the copy then has
$\rho_{\gamma'}/\rho_\gamma=\xi^4$ and $R'=R$, and the neutrinos supply exactly the free-streaming excess.
This statement does not provide a mechanism. Charged particles cannot mix with their mirror partners without
violating gauge invariance, and converting photons alone (a ``hidden CMB''~\cite{Jaeckel2008,Adams2026})
lowers the baryon-to-photon ratio at BBN by $\lambda^{-3/2}$, which raises D/H by about $20\%$, and yields
free-streaming dark photons, which lie $+13.7$ above \LCDM{} even after a refit (Sec.~\ref{sec:sym_demands}). In
this construction the mirror plasma and the extra free-streaming radiation therefore have to be created after
BBN; they cannot be converted from the visible plasma.

Late production of dark radiation has a long
history~\cite{Fischler2011,Hooper2012,Aloni2023}, and for free-streaming radiation from a decaying
particle the limits are strong: the particle may carry at most about $3\%$ of the energy
density~\cite{Sobotka2024}. Here a relic $X$ must carry $(\lambda^2-1)/\lambda^2\approx14\%$ when it
decays, which is admissible only because its products form an acoustic mirror plasma and free-streaming
radiation in the proportions that the symmetry requires.

\subsection{The open problem: mirror baryons}\label{sec:mirror_baryons}
The requirement that we cannot meet is the mirror baryon content. The symmetry needs the baryon loading
of the mirror plasma to equal the visible one. Since $R'/R=(m_p'/m_p)(\eta'/\eta)/\xi$, where $\eta$ and
$\eta'$ are the baryon-to-photon ratios, this means
\begin{equation}
\frac{\eta'}{\eta}=\xi\,\frac{m_p}{m_p'} .
\label{eq:etaprime}
\end{equation}
For synchronous dark constants $\eta'=\eta$. For the exact copy, $\eta'=\xi\eta=0.63\,\eta$ at
$\lambda=1.077$: the mirror asymmetry must be \emph{smaller} than the visible one by the temperature
ratio, which no symmetry suggests. In the code the condition is imposed through
$\omega_b'=(\lambda^2-1)\,\omega_b$ in both cases, and the refits tolerate $R'/R$ from below $0.55$, the lowest value refitted, to $1.03$, a range that includes the matched value (Table~\ref{tab:requirements}).

A symmetric cascade creates no net baryon number. The asymmetry must come either from CP-violating
decays of $X$ with the yield of Eq.~\eqref{eq:etaprime}, at photon temperatures of $16$ to
$1.6$\,keV for the lifetimes of Sec.~\ref{sec:therm}, or from a
pre-existing asymmetric dark matter population~\cite{Kaplan2009,Petraki2013,Zurek2014} that the decay
heats and re-ionizes, in which case Eq.~\eqref{eq:etaprime} is a coincidence. Mirror baryogenesis does
not give it automatically~\cite{AlonsoAlvarez2024,Bittar2023}. We have no mechanism. There is a second
coincidence: the energy of the relic must be close to $(\lambda^2-1)(\rho_\gamma+\rho_\nu)$ for the $\lambda$ set by $\kappa$ (the refits prefer about $20\%$ less, with a width of $\pm16\%$; Sec.~\ref{sec:requirements}), while in our constructions the abundance of $X$ and the
coupling $\kappa$ are independent.

\subsection{Requirements from the likelihood}\label{sec:requirements}
We measured what else the production must deliver, at $\lambda=1.077$. These tests compare mirror
sectors at a fixed history of the constants, so they use fits with prescribed constants, as in Part~I,
with a tabulated one-field history (the score does not depend on how the constants return,
Sec.~\ref{sec:symmetry}). A check with both fields evolved in the code agrees: moving the exact copy from $F=0.409$ to $F=0.478$ changes $S$ by $+0.2$, against $+0.5$ with prescribed constants. Table~\ref{tab:requirements}
separates single evaluations at fixed cosmology, which overstate the requirement, from refits; only the
refitted column should be read as a requirement. Each test moves along one path in the space of mirror
sectors, and the quantities are not independent. The temperature test changes the whole mirror sector
together, at $R'/R=1$ and $F=0.409$. The loading test changes the mirror baryon density alone. The
composition test moves energy between the free-streaming radiation and the mirror photons at fixed total
and keeps $R'/R=1$, so it also changes the mirror temperature and baryon density
(Sec.~S1 of the Supplemental Material lists the inputs). The widths in the refitted column are widths along these
paths for a change of $1$ in the score, not a joint confidence region.

\begin{table}[t]
\caption{Requirements on producing the mirror sector after BBN. ``Fixed'': tolerance for a change of
$1$ in the score at fixed cosmological parameters. ``Refitted'': the same after refitting the six
parameters, from the refits described in the text. For the temperature and for $R'/R$ the entry is the minimum and half-width of a parabola through the lowest refit and its two neighbors; for $F$ it is the same for a parabola through the four refits, whose lower edge, $F=0.24$, lies below the lowest refit, $0.34$; for $z_p$ it is interpolated between the computed points at $10^5$ and $3\times10^5$.}
\label{tab:requirements}
\begin{ruledtabular}
\begin{tabular}{lll}
Quantity & Fixed & Refitted\\
\colrule
Mirror temperature $\delta\xi/\xi$ & $-0.7\pm1.2\%$ & $-5.2\pm4.0\%$\\
Baryon loading $R'/R$ & $1\pm0.09$ & $0.76\pm0.27$\\
Free-streaming fraction $F$ & $0.409\pm0.025$ & $0.38\pm0.14$\\
Production redshift $z_p$ & $\gtrsim2\times10^{5}$ & $\gtrsim2\times10^{5}$\\
Dark constants & -- & exact copy allowed\\
\end{tabular}
\end{ruledtabular}
\end{table}

\begin{figure*}[t]
\includegraphics[width=\textwidth]{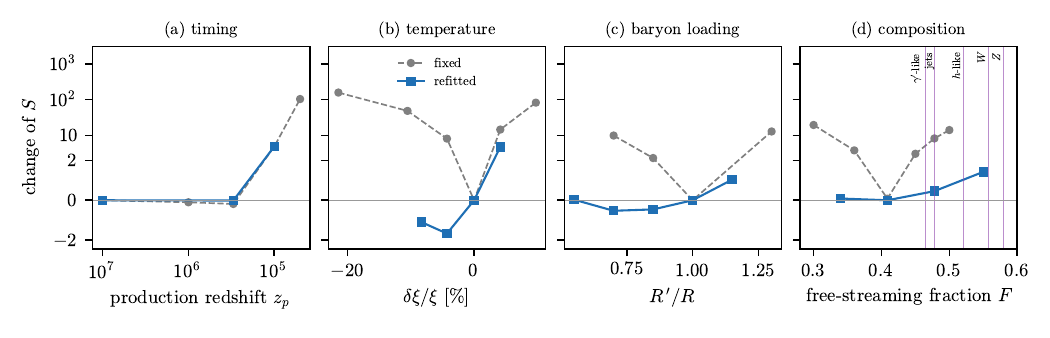}
\caption{Change of the score for departures from the symmetric mirror sector at $\lambda=1.077$, at
fixed cosmology (gray) and after refitting (blue). (a)~Production redshift of the mirror sector.
(b)~Mirror temperature relative to the value set by the kick. (c)~Baryon loading. (d)~Free-streaming
fraction $F$ of the mirror radiation; vertical lines mark decay channels of a mirror-sector relic
(dark-photon-like, quark jets, Higgs-like, $W'$ and $Z'$; Sec.~S4 of the Supplemental Material). All vertical axes are linear between $-2$ and $2$ and logarithmic above.}
\label{fig:production}
\end{figure*}

We first hold the constants at the values for $\lambda_c=1.077$ and move the mirror sector
to $\lambda_m\neq\lambda_c$. At fixed cosmology the cost grows quickly: $+8.1$ and $+14.5$ for
$\delta\xi/\xi=-4.3\%$ and $+4.1\%$. After refitting it is $-1.7$ and $+4.8$
(Fig.~\ref{fig:production}b). A fourth refit, at $\delta\xi/\xi=-8.3\%$, gives $-1.1$. The lowest refit, at $-4.3\%$, now has higher ones on both sides: a parabola through it and its two neighbors has its minimum at $\delta\xi/\xi=-5.2\%$, $1.7$ below the matched point, and rises by $1$ within $\pm4.0\%$, which corresponds to $\pm16\%$ in mirror energy density; the matched temperature lies just outside that range. The fitted
$H_0$ follows the mirror radiation and not the constants: it is $72.1$, $72.9$ and $74.8\kms$ in
the three refits away from the matched temperature, from the coolest mirror sector to the warmest.

For the baryon loading, at fixed cosmology $R'/R=0.85$ and $1.3$ cost $+2.3$ and $+12.9$.
Refitting at $R'/R=0.55, 0.70, 0.85$ and $1.15$ gives $+0.03, -0.53, -0.46$ and $+1.04$. The lowest refit, at $R'/R=0.70$, has higher ones on both sides; a parabola through it and its two neighbors has its minimum at $R'/R=0.76$, $0.6$ below the matched point, and rises by $1$ within $\pm0.27$.

The symmetry requires extra free-streaming radiation in a fixed proportion to the
mirror photons,
\begin{equation}
F\equiv\frac{\rho_{\rm fs}}{\rho_{\gamma'}+\rho_{\rm fs}}=\frac{0.691}{1.691}=0.409 ,
\end{equation}
where $0.691=N_{\rm eff}\tfrac78(4/11)^{4/3}$. At fixed cosmology the cost is
$\simeq1660\,(\delta F)^2$, and putting all of this energy into mirror photons costs
$+316$. Refits at $F=0.34$, $0.478$ and $0.55$ give $+0.1$, $+0.5$ and $+1.4$
(Fig.~\ref{fig:production}d). Along this path the mirror temperature follows the composition: at
$F=0.478$ it is $3.1\%$ lower and the mirror baryon density $12\%$ lower than at
$F=0.409$, so the cost of $+0.5$ is the cost of the whole path and must not be added to the
temperature tolerance above.

In a generic model this division of the decay energy would be a separate tuning. It is not if $X$
belongs to the mirror sector and decays into a mirror copy of the Standard Model: if every unstable
product decays before it loses energy, the share of the energy that ends in mirror neutrinos is fixed by
Standard Model branching ratios~\cite{PDG2024} and the primary channel. Such vacuum values are $F\approx0.48$ for
hadronic channels, which the refit at $F=0.478$ shows to cost only $+0.5$, and near $0.55$ for
weak-boson-like channels ($+1.4$). They are upper limits: muons with a Lorentz factor above about
$40$ lose their energy to a mirror photon bath of $5$\,keV before they decay, and if every muon
stops the fractions fall to $0.17$ and $0.21$--$0.33$ (Sec.~S4 of the Supplemental Material).

We model the timing of the production as the decay of a relic with rate $\Gamma_X=H(z_p)$, that is, with
lifetime $\tau_X=1/H(z_p)$, twice the age of the Universe at $z_p$ in the radiation era, into a
perfect-fluid component, which stands in for the tightly coupled mirror plasma while its baryon loading
is small, and a free-streaming component, in the proportions $1-F$ and $F$; both come from the same
decay. Relative to production at $z_p=10^7$, the cost at fixed cosmology is $-0.2$,
$+4.9$ and $+103$ for $z_p=3\times10^5$, $10^5$ and $5\times10^4$
(Fig.~\ref{fig:production}a). With the reference refitted too, these costs become $0.0$ and $+4.9$ for $z_p=3\times10^5$ and $10^5$. The threshold therefore lies between these two computed points; a logarithmic interpolation puts a cost of $1$ at $z_p\approx2\times10^{5}$.

Energy injected into the visible sector after
BBN is limited by photodissociation~\cite{Kawasaki2018} and by the COBE/FIRAS bound on $\mu$-type spectral
distortions~\cite{Fixsen1996,Hu1993,Chluba2012}, which require a visible branching ratio
$\lesssim10^{-7}$--$10^{-4}$ over the allowed lifetimes; a mirror-sector relic whose only portal is a
small kinetic or mass mixing $\epsilon_{\rm mix}\lesssim3\times10^{-4}$ satisfies this.

\subsection{Thermalization, and the relic during BBN}\label{sec:therm}
A blackbody mirror plasma requires processes that change the photon number,
chiefly double Compton scattering. For an exact-copy mirror plasma of hydrogen the rate of the
small-distortion estimate~\cite{Hu1993prd,Chluba2012} is $0.10$ of the visible one at a given redshift
(Sec.~S8 of the Supplemental Material), so a blackbody is reached only if the energy is released at
$z\gtrsim5\times10^{6}$, that is, for $\tau_X\lesssim10^{6}$\,s or $z_p\gtrsim7\times10^{6}$. The
estimate assumes a small distortion of an existing blackbody, while the mirror sector starts empty, so the
bound is indicative. For later production the mirror photons keep a deficit in number; their energy
density, which is what the symmetry constrains, is unaffected, but mirror recombination is delayed, a case
that we have not computed.

Before it decays the relic behaves as matter: for $\tau_X=10^4$\,s it carries
$2.6\%$ of the radiation density when deuterium forms. Added to the one-field kick of
Sec.~\ref{sec:bbnkick}, relics with $\tau_X=10^4$, $10^5$ and $10^6$\,s give $\chi^2=5.8$,
$3.7$ and $3.8$ with PRIMAT rates, against $4.0$ without the relic
(Sec.~S8 of the Supplemental Material gives the yields). If we ask that the fit be no worse than standard BBN at the
same baryon density ($\chi^2=6.1$), BBN allows lifetimes down to about $10^4$\,s, where helium is
$2.3\sigma$ high and the relic matters as much as the kick. These runs let the relic decay all at once at $t=\tau_X$. With an exponential decay and its radiation followed, the relic that gives the same final mirror radiation is $13\%$ more abundant; for $\tau_X=10^4$\,s it carries $2.9\%$ of the radiation density when deuterium forms, D/H changes by $+3.8\%$ and $\chi^2=6.4$, just above standard BBN, which moves the lower edge slightly above $10^4$\,s; the mirror radiation already released by then is $0.05\%$ of the radiation density.

Three separate requirements bound the lifetime: BBN from below, at about $10^4$\,s;
the thermalization estimate from above, at about $10^{6}$\,s; and the CMB fits from above, at
$\tau_X\lesssim8\times10^{8}$\,s ($z_p\gtrsim2\times10^{5}$). The first two give
$10^4\,{\rm s}\lesssim\tau_X\lesssim10^{6}$\,s, which corresponds to $z_p\approx7\times10^{7}$ to
$7\times10^{6}$ and to photon temperatures of $16$ to $1.6$\,keV at $z_p$. This window is the
intersection of requirements that were derived separately, each with its own approximations: the BBN
runs use the one-field kick and no mirror radiation, the thermalization bound is an estimate, and the
timing test uses a fluid proxy. We have not carried one model, with a single set of parameters, through
all three calculations. The closest we come is the two-field, exact-copy fit with $F=0.478$
(Table~\ref{tab:krm_fits}): BBN with the two-field kick and a relic of $\tau_X=10^5$\,s at the baryon
density of that fit gives $Y_p=0.2477$ and $\mathrm{D/H}=2.483\times10^{-5}$
($\chi^2=3.2$), with a helium fraction $0.0003$ above the one used in the fit. For lifetimes in the window the decay is complete before the modes that the CMB
data probe enter the horizon, which is why the fits with a mirror sector present from the start apply
(the cost of $z_p=10^6$ relative to $10^7$ is $-0.1$).

%=====================================================================================================
\section{Discussion}\label{sec:discussion}
%=====================================================================================================

\subsection{Relation to previous work}\label{sec:previous}
The scaling symmetry, its mirror-sector realization and the variation
of the constants at fixed binding energy are those of
Refs.~\cite{CyrRacine2022,Zhang2023,Greene2023,Greene2024}; we extend their tests to ACT DR6 and DESI DR2.
The cost of replacing the acoustic dark plasma was shown at the level of the spectra in
Ref.~\cite{Ge2023}, and Ref.~\cite{Greene2024} found the preference for earlier mirror recombination that
our exact-copy fits share, up to the untested recombination model of the dark sector
(Sec.~\ref{sec:sym_profile_text}). Section~\ref{sec:bbn} quantifies the nucleosynthesis constraint anticipated in
Refs.~\cite{CyrRacine2022,Zhang2023,Schoneberg2026b}, following earlier varying-constant
work~\cite{Coc2007,Dent2007,Clara2020,Deal2021,Martins2021,Seto2023,Seto2023me,Garramone2026}. The
ingredients of the kick are known (Sec.~\ref{sec:krm_prior}), our axion-like EDE fit reproduces
Ref.~\cite{Poulin2026}, and stellar ages have been used against the symmetry~\cite{Ge2023} and against
early-time solutions in general~\cite{Bernal2021,BoylanKolchin2021,Vagnozzi2022}.

Without a mirror sector, a varying $m_e$ or a modified recombination history, alone or with curvature or dynamical dark
energy~\cite{Hart2018,Hart2020,Sekiguchi2021,Lee2023,Schoneberg2025,Lynch2024,Lynch2024desi,Baryakhtar2024,Toda2026,Wang2026,Smith2025cpl,Lee2026,Calabrese2025},
generally reaches $H_0\approx69$--$71\kms$, because raising $H_0$ through recombination alone lowers
$\Omega_m$~\cite{Lee2026}; varying $\alpha$ alone reaches only $H_0\approx68.6\kms$~\cite{Toda2025alpha}, and an ultralight
scalar with quadratic couplings leaves the inferred $H_0$ unchanged~\cite{Ghosh2026}. In the symmetric
model the shift is tied to the dark radiation, and $\Omega_m$ is preserved to $1\%$. Triggered, coupled or
neutrino-based alternatives to axion-like EDE are studied in
Refs.~\cite{Karwal2022,Sakstein2020,CarrilloGonzalez2021,Baidya2026,Chowdhury2026}. Twin and mirror
sectors~\cite{Berezhiani2005,Chacko2006,Foot2014,Chacko2017,Chacko2018,Bansal2022} and atomic dark
matter~\cite{Kaplan2010,CyrRacine2013,Bansal2023,Hughes2024,Barron2026} are well studied, and models in
which dark radiation is coupled to part of the dark matter~\cite{BuenAbad2025} are strongly restricted by
ACT DR6~\cite{BuenAbad2026,Cvetko2025}; the symmetric model differs in scaling all densities together.

The scaling symmetry and its mirror-sector realization, the axion-like EDE fits, which reproduce Ref.~\cite{Poulin2026}
and serve here as a control, the kick of a scalar at a mass threshold as such, and the use of stellar ages against the symmetry
are not new. What we add are the test of TCEDE against the full likelihood, the test of the scaling symmetry with ACT DR6 and
DESI DR2, the size and the sign of the shift of $\alpha$ that a mirror sector present at BBN would need, and the use of the
annihilation kick to generate the shift at recombination.

\subsection{Limitations}\label{sec:limitations}
All results are optimizer-based best fits and profiles. The two fits at $\lambda=1.077$, restarted from the scaling image of the \LCDM{} best fit, end within $0.20$ of the originals, and neither is lower by more than $0.01$; a second start of the EDE fit ends about $5$ higher. No global search was attempted. The two pipelines differ in CLASS version, neutrino
treatment and non-linear prescription. For a control model fitted in both, $\DS$ agrees to $0.22$ when the lensing prescription is the same and to $0.24$ otherwise. The TCEDE and EDE fits use linear lensing power except in the HMcode refits; the small TCEDE gain without SH0ES shrinks in them from $3.2$ to $1.4$, while most of the EDE gain survives ($6.8$ of $8.7$).

The dark-atom pipeline rests on our corrections and additions to \texttt{class\_adm} (Sec.~S1 of the Supplemental Material), which were tested
through limits and conservation laws, and not against an independent code. Its non-linear prescription is flagged as unreliable
for interacting dark matter, and it computes the lensing potential over a reduced range of wavenumbers, which changes $\DS$ by at
most $0.2$ (Sec.~\ref{sec:pipelines}). The mirror sector is hydrogen-only in the main fits and uses the simpler
recombination model. For visible hydrogen, the fast recombination mode was checked against the full radiative-transfer
calculation at $\lambda=1.077$ only (Sec.~S1 of the Supplemental Material). The fixed-$H_0$ fits below the best fits, like the other EDE fits, were closed by independent optimizer runs alone (the last lowered the score by at most $0.06$); four earlier runs, two of them for the EDE best fit and its fit at $71\kms$, had met that criterion and still stood $0.05$ to $0.56$ above a point found later, so such fits can lie that far above their minima. Where the field is switched off ($H_0=68\kms$, and $67\kms$ for TCEDE), two evaluations of EDE with a small field scored higher, which does not exclude a lower minimum there. The Planck and ACT calibrations are tied, and an independent Planck
prior changes $\DS$ of the fits refitted in both conventions by between $-0.52$ and $-0.12$ (Sec.~\ref{sec:calib}).

The BBN treatment scales the charged-particle rates with the Gamow factor and one power of $\alpha$; it neglects finite-radius
corrections, the $\alpha$ dependence of the strong part of the nuclear binding energies, and resonances. In the second part of the
paper, the microphysical parameters are fixed at benchmark values chosen to keep the energy of the fields small, and the fits
depend on them.

\subsection{Untested aspects}\label{sec:untested}
We have not included SPT-3G, galaxy full-shape, Lyman-$\alpha$ or other small-scale data that could test the dark acoustic
oscillations of the $\approx3\%$ dark-atom component; combined with ACT, Planck and DESI BAO data, SPT-3G data mildly prefer modified-recombination
models, while the CMB data alone show no evidence for them~\cite{Camphuis2026}. We did not repeat the analysis with the Planck PR4 likelihoods. Whether the residual symmetry breaking
in recombination can be detected with future damping-tail data~\cite{Smith2025damping}, and whether the dark-atom component
leaves a detectable acoustic signature on small scales, are questions that our fits do not answer.

The results of Part~II are conditional benchmarks. The kick, the damped fields, the production of the mirror sector and its
thermalization were each tested separately, at chosen values of the mass, the damping rate, the switch scale, the composition and
the relic abundance, and no single model with one set of parameters was carried through BBN, thermalization and the CMB. The
damping term is phenomenological, the initial conditions of the fields ($\varphi=0$, $\sigma=v_\sigma$) are assumed, the electron
source after BBN and the energy exchanged with electrons and baryons when their masses relax, of order $10^{-5}$ and $10^{-6}$ of
the total, are not included in the Boltzmann code, and the model is radiatively unstable (Sec.~\ref{sec:krm_prior}). A potential
steeper than quadratic, which would store less energy, was not computed. The thermalization of the mirror plasma and the energy
losses of the cascade particles are estimated only, the timing test uses a fluid proxy for the mirror plasma, and the cascade
fractions are semi-analytic vacuum values (Secs.~\ref{sec:requirements} and~\ref{sec:therm}). The origin of the mirror baryon
asymmetry and of the link between the relic abundance and the kick, a partial overlap of the decay with nucleosynthesis, and the
mirror helium abundance left by the cascade, which we take to be zero, have not been addressed.

%=====================================================================================================
\section{Conclusions}\label{sec:conclusions}\label{sec:newresults}
%=====================================================================================================

\begin{table*}[t]
\caption{The models compared. $\DS$ is relative to \LCDM{} in the same pipeline with the same supernova
data, and ``extra'' counts the parameters varied in the fit, with $\lambda$, which is scanned and not fitted, counted as one. $H_0$, in $\mathrm{km\,s^{-1}\,Mpc^{-1}}$, is that of the fit without SH0ES. TCEDE and EDE use linear lensing power except in the HMcode rows (the 2020 version); the dark-atom fits use the 2015 version of HMcode. The last three columns state whether the
model as fitted to the CMB also passes nucleosynthesis and equivalence-principle tests, and whether its
ingredients have a known origin; ``standard'' means that BBN is unchanged, so the yields are those of
standard BBN at the fitted $\omega_b$. With SH0ES, TCEDE and EDE are refitted, and their best fits move to $H_0=69.9$ and
$72.4\kms$ (Table~\ref{tab:tcede_free}). The dark-atom and modulus rows evaluate the fits without SH0ES, so their entries with SH0ES are upper bounds on the minimized score (Tables~\ref{tab:sym_profile} and~\ref{tab:krm_fits}), except the two entries at $\lambda=1.077$, which direct fits to the SH0ES data confirm; the table footnotes mark which is which. For axion-like EDE the potential is put in by
hand; for TCEDE the potential follows from the coupling to the neutrino (``potential only''), and in both models the epoch of the transient is set by a free parameter, which in TCEDE is the initial value of the field~\cite{Kamionkowski2025}. The modulus rows include the energy of the field(s); ``kick only'' means that the shift of the
constants has an origin and the mirror sector does not. In the modulus rows ``yes'' under BBN refers to the
kick without the relic of Sec.~\ref{sec:therm}, and ``yes'' under the equivalence principle assumes that both
fields are damped and take their cosmological values near the Earth (Sec.~\ref{sec:ep}). In all
scaling-symmetry rows the single scanned
direction is $\lambda$. The scaling symmetry fixes the mirror baryon loading ($R'=R$) and the free-streaming fraction ($F=0.409$). The main fits assume a hydrogen-only mirror sector. Section~\ref{sec:sym_demands} tests the effect of adding mirror helium. In the modulus rows the mass, the damping rate, the switch scale, the initial field values and the
abundance of the mirror sector are benchmark choices, not fitted parameters (Part II).}
\label{tab:compare}
\begin{ruledtabular}
\begin{tabular}{lccccccc}
Model & Extra & $H_0$ & $\DS$ & $\DS$ (SH0ES) & BBN & Equivalence & Origin of\\
 & & without & & & & principle & ingredients\\
 & & SH0ES & & & & & \\
\colrule
TCEDE & 2 & 69.2 & $-3.2$ & $-14.3$ & standard & yes & potential only\\
\quad HMcode lensing, refitted & 2 & 69.2 & $-1.4$ & -- & standard & yes & potential only\\
Axion-like EDE & 3 & 71.0 & $-8.7$ & $-34.7$ & standard & yes & no\\
\quad HMcode lensing, refitted & 3 & 70.9 & $-6.8$ & -- & standard & yes & no\\
Scaling symmetry, synchronous & 1 & 71.9--75.2 & $-1.3$ to $-1.8$ & $-27.5$\footnotemark[1] to $-31.4$\footnotemark[2] & if after BBN & not addressed & no\\
\quad exact-copy dark atoms & 1 & 71.9--75.1 & $-2.8$ to $-3.7$ & $-29.1$\footnotemark[1] to $-33.0$\footnotemark[2] & if after BBN & not addressed & no\\
\quad with one-field modulus & 1 & 73.6 & $+3.1$ & $-26.6$\footnotemark[1] & yes & no & kick only\\
\quad with two-field modulus & 1 & 73.7 & $+1.4$ & $-28.3$\footnotemark[1] & yes & yes & kick only\\
\quad two-field modulus, exact copy & 1 & 73.8 & $-0.7$ & $-30.4$\footnotemark[1] & yes & yes & kick only\\
\end{tabular}
\end{ruledtabular}
\footnotetext[1]{Upper bound on the minimized score: the fit without SH0ES, evaluated with the SH0ES data.}
\footnotetext[2]{Fits to the SH0ES data at $\lambda=1.077$ end within $0.01$ of these entries.}
\end{table*}

We have compared two ways to raise the Hubble constant inferred from CMB data, an injection of energy before recombination and a
scaling symmetry of the cosmological observables, using the same Planck, ACT DR6, CMB lensing, DESI DR2 and Pantheon+ likelihoods.
Table~\ref{tab:compare} summarizes the models.

TCEDE is a complete field-theory model in which the coupling to the neutrino supplies the potential, tested here, to our knowledge, for the first time against the full
likelihood. We find that it raises $H_0$ only to $\approx69.2\kms$, with a gain that is within the uncertainty of the non-linear lensing model, and that it cannot reach $71\kms$ without a
penalty of $16$ with respect to \LCDM{}, which comes from the Planck polarization spectra at $\ell\approx200$--$400$. Axion-like
EDE, fitted in the same pipeline, improves on \LCDM{} by $8.7$ at $H_0=71.0\kms$ ($6.8$ with HMcode), and at
$71\kms$ it lies at $\DS=-8.7$ relative to the \LCDM{} best fit, where TCEDE lies at $+16$. The shortfall is therefore
specific to TCEDE. The potential of the axion-like model, however, is
put in by hand, and in both models the onset near matter--radiation equality is set by a free parameter.

The scaling-symmetry direction survives ACT DR6 and DESI DR2. A mirror sector with dark atoms, together with a Rydberg-preserving
shift of $\alpha$ and $m_e$, leaves these data nearly indifferent to $H_0$ once the cosmological parameters are refitted (the
scaling image of the \LCDM{} best fit is $54$ worse than \LCDM{} before the refit). Without SH0ES the fits lie
$1.3$--$3.7$ below \LCDM{}, which is not a preference, and with SH0ES they lie $27.5$--$33$ below it. This is what
Ref.~\cite{Greene2024} found on earlier data, so that the new data have not closed the direction. Without SH0ES, axion-like EDE, with two more parameters and an onset put in by hand, fits better than the symmetry at $72\kms$ and worse at $75\kms$; the fits do not locate the crossing (Fig.~\ref{fig:profiles}). The two substitutes
for the acoustic dark plasma that we tested fail, and the data accept a mirror sector with the visible sector's constants. The
symmetry is limited by absolute time scales: at $\lambda=1.077$ the Universe is $12.75$\,Gyr old, $2.5\sigma$ younger than
the oldest globular clusters in the most recent determination with errors added in quadrature, and $3.1\sigma$ with their
formation delay, and a one-sided limit at $95\%$ confidence gives $H_0\lesssim72\kms$ along the direction
(Sec.~\ref{sec:clocks}).

A thermal mirror sector present at nucleosynthesis is excluded ($\chi^2=159$) unless $\alpha$ was lower at that epoch, by
$2.4$--$3.7\%$ for our baseline inputs and by up to $5.7\%$ for the others that we considered, opposite in sign to the
shift at recombination. The sign holds for every choice of nuclear rates, helium data and baryon density that we considered, and a
single light scalar supplies such a history only with tuning. This motivates, but does not require, a sector that appears after
BBN: a history of the constants that changes sign, a sector only partly populated at BBN, or mirror baryons that precede most of
the mirror radiation remain possible.

A light scalar that is kicked at $e^\pm$ annihilation produces the shift of the constants at recombination with the right sign, and
ties its size to one coupling; $H_0$ follows only if the mirror radiation is produced with the matching abundance. With the energy of
the field in the calculation, the fit at the benchmark is worse than \LCDM{} without SH0ES ($\DS=+3.1$), while most of the gain with SH0ES remains ($\DS=-27$). One field violates the equivalence principle by seven orders of
magnitude. A second field that switches the coupling off cures this, is kicked as well, and for a small switch scale stores less
energy. With both fields damped, at a rate we do not explain, the two-field benchmark passes the equivalence-principle test and
fits the data better than the one-field model: $\DS=+1.4$ without SH0ES, and $-0.7$, as good as \LCDM{}, with exact-copy dark atoms ($-28$ and $-30$ with SH0ES). Creating the mirror sector after BBN requires a relic with the right abundance and decay products, and a mirror baryon
asymmetry tied to the visible one by Eq.~\eqref{eq:etaprime}, for which we have no mechanism.

We conclude that the present CMB, BAO and supernova data do not by themselves exclude a Hubble constant of $72$--$75\kms$ along
the scaling symmetry. Axion-like EDE, whose potential is put in by hand, returns to the \LCDM{} score at
$H_0\approx74\kms$, while TCEDE, whose potential follows from its coupling, reaches $71\kms$ only at a cost of $16$. For the
symmetry a complete realization is missing. A thermal mirror sector present at BBN would need a lower $\alpha$ at that epoch, and
one produced later needs a mirror baryon asymmetry for which we have no mechanism. The kick that we studied violates the equivalence
principle unless its coupling is switched off, which we achieved with a second field and a damping that we do not explain. The
Universe at $\lambda=1.077$ is also younger than the oldest globular clusters, as quoted above. The scaling symmetry is therefore a
direction that the data leave open, and not yet a solution of the Hubble tension. Improved primordial abundances, the CMB damping
tail and the ages of the oldest stars will test it further.

\emph{Code and data availability.} The changes to \texttt{class\_adm} and to CLASS, the likelihood settings, the search ranges and starting points of the fits, the score of each likelihood at each fit, the inputs of the BBN runs and the points behind Figs.~\ref{fig:profiles} and~\ref{fig:production} are provided with the submission as a code and data archive. For the TCEDE fits at fixed $H_0=67$ to $69.5\kms$, the axion-like EDE fits at $68$ to $70\kms$ and the proxy-plasma fits, the archive contains the total scores only. A public repository will be linked in the published version.

\begin{acknowledgments}
We acknowledge the use of the public likelihoods of Planck, ACT and DESI, of the
Pantheon+ and SH0ES data products, and of the codes CLASS, \texttt{class\_adm}, Cobaya, PRyMordial,
HyRec and Py-BOBYQA. AC thanks the Mercor Research team for encouraging a comprehensive study of
Hubble tension using frontier AI models. We acknowledge the use of Claude/Fable 5.1 and 5.2 models of Anthropic for code development,
testing, and the numerical runs, and for assistance in preparing the manuscript. The authors accept all responsibility for calculations and
the content of this paper.
\end{acknowledgments}

%=====================================================================================================
%=====================================================================================================

\clearpage
\setcounter{section}{0}
\setcounter{equation}{0}
\setcounter{table}{0}
\setcounter{figure}{0}
\renewcommand{\thesection}{S\arabic{section}}
\renewcommand{\theequation}{S\arabic{equation}}
\renewcommand{\thetable}{S\arabic{table}}
\renewcommand{\thefigure}{S\arabic{figure}}
\section*{Supplemental Material: Energy injection versus scaling symmetry as solutions to the Hubble tension after ACT DR6 and DESI DR2}

\noindent Sections, equations, tables and figures labeled with an S belong to this Supplemental Material; all other
numbers refer to the paper, whose notation is used throughout.

\section{Changes to \texttt{class\_adm}}\label{app:code}
Table~\ref{tab:pipelines} summarizes the two pipelines of Sec.~II\,B.
The dark-atom pipeline uses the public code \texttt{class\_adm}~\cite{Bansal2023}, in the version of its
repository\footnote{\url{https://github.com/jp-barron/class_adm-3.1}} identified by the git commit \texttt{c700a815}.
The TCEDE pipeline uses the public code CLASS~\cite{Lesgourgues2011,Blas2011}\footnote{\url{https://github.com/lesgourg/class_public}}
in its version 3.3.4, as contained in the repository of
DRMD-CLASS~\cite{Garny2026drmd}\footnote{\url{https://github.com/NEDE-Cosmo/DRMD-CLASS}} at the git commit
\texttt{aa2b61a0}. The model that DRMD-CLASS adds to CLASS is switched off in all our runs. The thermo-coupled model and
the axion-like potential are our own additions to this code (Sec.~\ref{app:tcede}).
Both pipelines evaluate the likelihoods with
Cobaya~\cite{Torrado2021}\footnote{\url{https://github.com/CobayaSampler/cobaya}} (version 3.5.7) and the packages
\texttt{act\_dr6\_cmbonly}\footnote{\url{https://github.com/ACTCollaboration/DR6-ACT-lite}} (version 1.0.0) and
\texttt{act\_dr6\_lenslike}\footnote{\url{https://github.com/ACTCollaboration/act_dr6_lenslike}} (version 1.2.1), and
minimize with Py-BOBYQA~\cite{Cartis2019}\footnote{\url{https://github.com/numericalalgorithmsgroup/pybobyqa}} (version 1.5.0).
The dark-atom pipeline uses HyRec~\cite{Lee2020}\footnote{\url{https://github.com/nanoomlee/HYREC-2}} as contained in
\texttt{class\_adm}. Our changes to \texttt{class\_adm}, to CLASS for the TCEDE pipeline and to the nucleosynthesis code of
Sec.~\ref{app:bbn} are in the archive that accompanies the paper.

\begin{table*}[t]
\caption{Pipelines.}
\label{tab:pipelines}
\begin{ruledtabular}
\begin{tabular}{lp{0.19\textwidth}p{0.31\textwidth}p{0.27\textwidth}}
Pipeline & Models & Boltzmann code & Fits\\
\colrule
TCEDE & TCEDE, axion-like EDE, \LCDM & CLASS 3.3.4 with coupled scalar and neutrino, or axion-like scalar; linear lensing power & BOBYQA with polish; fixed-$H_0$ profiles\\
Dark-atom & scaling symmetry, \LCDM; Part II models & \texttt{class\_adm} (CLASS 3.1): dark atoms, varying constants; in Part II the modulus evolved in the code; HMcode & BOBYQA with polish at fixed $\lambda$\\
\end{tabular}
\end{ruledtabular}
\end{table*}

\emph{Parameters of the mirror sector.} With $\omega_{\rm dm}$ the total dark matter density including
the dark baryons, the inputs at scaling parameter $\lambda$ are: \texttt{r\_all\_twin}
$=\omega_b'/\omega_{\rm dm}$ with $\omega_b'=(\lambda^2-1)\,\omega_b$, which imposes $R'=R$;
\texttt{xi\_twin} $=\xi=(\lambda^2-1)^{1/4}$; \texttt{m\_p\_dark}, \texttt{m\_e\_dark} and
\texttt{alphafs\_dark}, the dark proton and electron masses and the dark fine-structure constant
(Sec.~IV); \texttt{YHe\_twin}, the mirror helium fraction, zero unless stated; and
$N_{\rm ur}=2.0308\,\lambda^2$, with the massive neutrino rescaled as in Sec.~IV. The requirement tests of Sec.~VII\,B change these inputs as follows. \emph{Temperature:}
every mirror input is set to its value for $\lambda_m$ ($N_{\rm ur}$, $\xi$, $\omega_b'$, the rescaled
neutrino and, for synchronous constants, the dark masses and coupling, which depend on $\xi$), while the
visible constants keep the history for $\lambda_c=1.077$; $R'/R=1$ and $F=0.409$ are unchanged. The
points are $\lambda_m=1.055$, $1.065$ and $1.090$, that is $\xi=0.580$, $0.605$ and $0.659$ against
$0.632$. \emph{Baryon loading:} $R'/R=r$ is set by $\omega_b'=r\,(\lambda^2-1)\,\omega_b$ at fixed $\xi$,
$F$ and dark constants, for $r=0.55$, $0.70$, $0.85$ and $1.15$; the cold dark matter is refitted.
\emph{Composition:} energy is moved between $N_{\rm ur}$ and the dark photons at fixed total, with
exact-copy dark constants, which do not change. The dark-photon density then scales as
$(1-F)/(1-0.409)$, the temperature $\xi$ as its fourth root, and $\omega_b'$ is rescaled by the same
factor as the dark-photon density, so that $R'/R$ stays $1$. For $F=0.34$, $0.478$ and $0.55$ this gives
$\xi=0.650$, $0.613$ and $0.591$, $\omega_b'/[(\lambda^2-1)\omega_b]=1.116$, $0.883$ and $0.761$, and
$N_{\rm ur}=2.274$, $2.438$ and $2.524$ against $2.356$. The evaluations at fixed cosmology in Fig.~5 follow the same rules, at $\lambda_m=$\,$1.03$, $1.05$, $1.065$, $1.09$ and $1.11$ (that is, $\delta\xi/\xi=$\,$-21.4$, $-10.5$, $-4.3$, $+4.1$ and $+9.8$ per cent), $R'/R=$\,$0.70$, $0.85$ and $1.30$, and $F=$\,$0.30$, $0.36$, $0.45$, $0.478$ and $0.50$.

\emph{Corrections.} (i)~When varying constants and the dark sector are both on, the temperature
rescaling in the HyRec interface was applied twice; it is now applied once. (ii)~The dark-sector drag
block of the thermodynamics module ran even without a dark sector and read an unset table column; it
is now guarded. (iii)~The recombination mode of visible hydrogen (footnote in Sec.~II): the
simpler mode is now used for the dark sector only. Without this correction, a mirror model with
physically negligible $\lambda-1$ refits to a score $1.6$ below \LCDM{} with $n_s=0.966$, so the error
biases both the score and the parameters. Corrections (i) and (ii) remove failures. Without (i) our runs stop with an error whenever a dark sector and
varied constants are both present, which is the case for every fit at $\lambda>1$: the dark sector leaves
its rate table at $z\approx14$. Correction (ii) matters only without a dark sector, where an unset
value could crash the run. Neither changes a score that the original code returns. Correction
(iii) is the one that restores the agreement with \LCDM{} to $0.04$ at $\lambda=1.001$, and it acts at
every $\lambda>1$. The public version of \texttt{class\_adm} does not contain these changes; they are provided as patches
against it, with the additions below.

\emph{Additions.} (iv)~Constants can be read from a table of $\alpha(z)$ and $m_e(z)$. (v)~A flag turns
the radiation produced by a decaying species into a perfect fluid, and the decay energy can be divided
between that fluid and a free-streaming component; this is used for the timing test of
Sec.~VII. (vi)~The modulus of Sec.~VI\,C: one or two fields $\varphi_i$
with masses $m_i$ and damping rates $\Gamma_i$, in units of the reduced Planck mass and conformal time,
\begin{align}
\varphi_i''+(2\mathcal H+a\Gamma_i)\varphi_i'+a^2m_i^2\varphi_i&=0,\nonumber\\
\rho_{\rm dr}'+4\mathcal H\rho_{\rm dr}&=\sum_i\Gamma_i\varphi_i'^2/a,
\end{align}
with the field energies and pressures and $\rho_{\rm dr}$ in the Friedmann equation. In synchronous
gauge the perturbations obey
\begin{equation}
\delta\varphi_i''+(2\mathcal H+a\Gamma_i)\delta\varphi_i'+(k^2+a^2m_i^2)\delta\varphi_i
+\tfrac12h'\varphi_i'=0 ,
\end{equation}
and the decay radiation follows the free-streaming hierarchy of a massless species, with the perturbed
energy transfer $2a\Gamma_i\varphi_i'\delta\varphi_i'/a^2$ in its monopole equation and the momentum
transfer $a\Gamma_i\,k^2\varphi_i'\delta\varphi_i/a^2$ in its dipole equation. These terms follow from
one covariant prescription: the friction acts in the rest frame of the field,
$\Box\varphi_i-m_i^2\varphi_i=\Gamma_i\,u^\mu\partial_\mu\varphi_i$ (signature $-{+}{+}{+}$) with $u^\mu$ the
future-directed unit normal to the surfaces of constant $\varphi_i$, so that the field loses energy and momentum at the rate
$Q_\nu=-\Gamma_i\,(u^\mu\partial_\mu\varphi_i)\,\partial_\nu\varphi_i$, which the decay radiation
receives. The total energy--momentum tensor is then conserved identically, for the background and for
the perturbations. The constants are
computed from the evolved fields. The fields start frozen at their post-kick values. The decay radiation is not part
of the closure relation, so the Hubble rate today exceeds the input $H_0$ by a fraction
$1\times10^{-5}$ for one field and $8\times10^{-6}$ for two.

\emph{Tests of the field component.} (a)~The decay radiation equals the integrated source to
$3\times10^{-6}$, and the field energy obeys its continuity equation to $10^{-5}$. (b)~With the field
amplitude scaled to zero at fixed shift of the constants, the spectra agree with those for prescribed
constants to $1\times10^{-5}$. (c)~Neither the field nor its decay radiation clusters on sub-horizon
scales. The resulting drop of $\sigma_8$ at fixed parameters, $0.92\%$ for one field and
$0.64\%$ for two, agrees with an independent integration of the growth equation in the same
background ($0.91\%$ and $0.63\%$). Setting the perturbations of the field to zero loses
this effect and is not a valid approximation in synchronous gauge. (d)~For an undamped field, the field blocks reproduce the scalar-field module of CLASS to $5\times10^{-9}$ in the background, to $4\times10^{-5}$ in TT, $2\times10^{-4}$ in EE and $4\times10^{-7}$ in the lensing power, and to $3\times10^{-8}$ in the rms matter fluctuation at a fixed radius. (e)~For the damped two-field benchmark, in a test build that prints the perturbations of the fields and of the decay radiation, the perturbed energy and momentum equations of the combined system hold with median residuals of $3\times10^{-9}$ and $4\times10^{-8}$ relative to the largest term, and 95th percentiles of at most $3\times10^{-5}$ (printed time steps at $z<1500$, $k=0.005$--$0.3\,{\rm Mpc}^{-1}$, time derivatives from cubic splines of the printed histories). With second-order differences in place of the splines the median residual of the momentum equation grows by a factor of $3.9$--$4.0$ for each doubling of the step, so what remains is the error of the numerical derivative. The fields alone miss by order unity (median $0.4$--$1.0$); this is the transfer that the decay radiation takes up.

\emph{Recombination at finite $\lambda$.} HyRec's \texttt{SWIFT} mode replaces the radiative-transfer
calculation by a correction function that was computed and tabulated, with its first derivatives in the cosmological
parameters, at the standard constants~\cite{Lee2020}. Its full mode cannot run inside CLASS, so we built
the standalone program from the same sources and ran both modes at the best-fit parameters, with the
expansion rate scaled through the matter density and the effective neutrino number and with the shifted
constants. For \LCDM{} the two modes agree to $7\times10^{-5}$ in the ionization fraction $x_e$ over
$500<z<1700$. At $\lambda=1.077$ the full calculation gives $x_e$ lower by up to $0.05\%$ near
$z=1340$ and higher by up to $0.16\%$ near $z=1010$. The shifted constants alone produce
$0.17\%$ and the scaled expansion alone $0.02\%$: when the atomic rates and the expansion rate
are scaled together the radiative transfer is nearly unchanged, while the correction function responds to
the expansion rate as if the rates were standard. (CLASS passes the matter density without the dark
baryons and an effective neutrino number evaluated before dark $e^\pm$ annihilation to that function,
which changes $x_e$ by $0.02\%$; the comparison uses the same inputs, and the standalone fast mode
then reproduces the $x_e$ of CLASS to $1\times10^{-4}$ through recombination.) We applied the ratio of the
two modes to the visible $x_e$ in CLASS. At fixed parameters, with the calibration re-profiled, the peak
of the visibility function moves by $-0.15$ in redshift and the score changes by $-0.003$ for
\LCDM{}, $+0.20$ for the synchronous fit and $+0.01$ for the exact copy.
Refitted with the correction, the two fits give $\DS=-1.57$ and $-3.37$, against $-1.67$ and $-3.30$ without it. The error of the fast mode is therefore below the $0.3$ that we treat as meaningful; it widens the difference between the two choices of dark constants by $0.2$. The dark sector itself uses the simpler mode without any radiative-transfer correction; we
did not test that choice.

\emph{Minimization.} After the restarts of Sec.~II\,C, every fit of the TCEDE pipeline was closed, in rounds except for the axion-like EDE fits, the TCEDE fit at $H_0=71\kms$ with the independent Planck calibration and the TCEDE best fit with HMcode. In each round a least-squares quadratic fitted to the evaluations within $3$ (where these were too few, $6$) of the lowest score of the fit's configuration (for a free-$H_0$ fit including those of its fixed-$H_0$ fits) proposed steps inside a trust region, and one evaluation was placed about one standard deviation from the best point on each side of every curvature direction that the existing evaluations did not yet reach; the HMcode and independent-calibration fits, which have few evaluations, take the curvature of the corresponding linear fit with tied calibration. A fit counts as closed at the first round that lowers its score by less than $0.03$. Each fit closed with a curvature taken from another fit, the TCEDE best fit, the TCEDE fit at $H_0=70\kms$ (one point for both supernova choices) and one fit of each remaining family were then each checked by an independent optimizer run from the closed point (first radius $0.4$ in scaled units, at most $45$ evaluations, then, where it gained $0.03$ or more, a line search along its net move). Over the evaluations that the closing uses, the scores scatter about the fitted quadratic by $0.015$ to $0.048$ (rms; the largest for the TCEDE fit at $H_0=70\kms$ with SH0ES) where the quadratic is fitted in full, and by $0.076$ for the TCEDE HMcode fit at $H_0=71\kms$, whose quadratic takes its curvature from the linear fit; because a round measures its gain against the lowest evaluated point, a closing can stop a few hundredths above the minimum of the smooth surface. A check therefore passes if it gains less than $0.1$, more than twice the scatter where the quadratic is fitted in full (the check of the HMcode fit at $H_0=71\kms$ gained $0.021$ at standard precision); after a gain of $0.03$ or more at standard precision the new point is also evaluated at higher precision, and the row takes the lower of the two higher-precision scores. The axion-like EDE fits do not suit a closing in rounds: closed in rounds within $1.5$ of its minimum, the free fit gained $0.12$ at higher precision in its check, and for the fit at $H_0=75\kms$ and the free fit with HMcode too few evaluations exist for a quadratic. The EDE fits were therefore closed by checks alone, each starting from the best point of the previous one, until one passed (at most three per fit); for the second minimum of the free fit (Sec.~III\,C) the checks were bounded to its basin, $\theta_i<1.6$. Of the fits in the tables, two that had passed, the fit at $H_0=71\kms$ and the free fit, were reopened when a later evaluation lay more than $0.05$ below them, and closed again in the same way from that point. The TCEDE fit at $H_0=71\kms$ with the independent Planck calibration and the TCEDE best fit with HMcode, closed in rounds with a curvature taken from another fit, each gained $0.13$ at higher precision in its check, and were closed by checks in the same way (two runs each). The largest gain of a passing check was $0.08$ (for the second minimum of the axion-like EDE fit, at higher precision). Every row rests on a check, of its own fit or of one fit of its family, that gained less than $0.1$ at higher precision (or less than $0.03$ at standard precision); for the other members of a family the check is indirect. The quoted score is the lowest higher-precision evaluation of the fit: at the closed point, at the check's best point, or at an earlier point where that one scores lower. Some fits move far along the direction in which $A_se^{-2\tau}$ and the calibration parameters compensate: those with an independent Planck calibration (Sec.~II\,D), those in another code (Sec.~IV\,C) and the refits of Sec.~VII\,B that bracket a minimum. A search over all parameters at once crosses that direction slowly: for \LCDM{} with the independent calibration it stopped $0.38$ above the minimum while meeting the restart rule of Sec.~II\,C. In these fits the calibration parameters are therefore minimized inside every evaluation, with the spectra held fixed, each restart is followed by a line search along its net move, and the fit stops when a restart gains less than $0.05$. In the tied convention, where the fits start close to their minimum along that direction, the two procedures agree to $0.03$ for \LCDM{}. Two fits at $\lambda=1.077$ were also restarted from the scaling image of the \LCDM{} best fit instead of a neighboring fit. The restart ends within $0.01$ of the original fit for synchronous dark constants and $0.20$ above the original fit for the exact copy.

After the TCEDE fits at fixed $H_0$ were found to stop short of their minima (Sec.~II\,C), every dark-atom fit quoted in the paper was tested twice. First, a quadratic in the eight fitted parameters was fitted to all evaluations of the fit and its minimum evaluated, with the calibration parameters minimized at fixed spectra, until the quadratic predicted a gain below $0.03$. This lowered nine of the $27$ fits by more than $0.05$ (at most $0.68$), but it can only find points inside the region that the stored evaluations cover: for the \LCDM{} fit to the SH0ES data it predicted $0.02$ where the minimum lay $0.39$ lower, and evaluations on a stencil along the principal axes of the quadratic did not find that point either. Second, each fit was therefore restarted from its lowest point with the procedure of the previous paragraph, and closed when a stage gained less than $0.05$. This lowered ten fits by more than $0.05$ (at most $0.35$). In all, 14 fits are replaced by lower points (production at $z_p=10^7$, $0.82$; production at $z_p=10^5$, $0.58$; production at $z_p=3\times10^5$, $0.49$; $\lambda=1.095$, $0.46$; one field, $\Gamma=m/8\pi$, $0.39$; mirror sector for $\lambda=1.065$, $0.30$; $F=0.340$, $0.28$; two fields, exact copy, $F=0.478$, $0.21$; two fields, $0.18$; $\lambda=1.050$, $0.18$; control, half an extra species, $0.08$; prescribed field history, $0.08$; mirror sector for $\lambda=1.090$, $0.07$; $\lambda=1.077$, exact copy, $0.06$); for the others no evaluated point lies more than $0.04$ below the fit.

\emph{Calibration convention.} Table~\ref{tab:calibfull} extends Table~I to all fits. In the rows
marked ``line'' the best fit of the tied convention is moved along one direction only, the displacement of the
\LCDM{} best fit between the two conventions, in steps of half that displacement, with the calibration parameters
re-profiled at each point; the entry is the lowest point and is an upper bound that lies closer to a refit than the
entry at fixed cosmology (for \LCDM{} itself, $0.58$ above the refit at fixed cosmology and
$0.11$ above it on the line).

\begin{table}[t]
\caption{$\DS$ in the two calibration conventions for all fits; the last column is explained in the text.}
\label{tab:calibfull}
\begin{ruledtabular}
\begin{tabular}{lccc}
Fit & tied & independent & cosmology\\
\colrule
\multicolumn{4}{l}{TCEDE pipeline (linear lensing power)}\\
TCEDE best fit & $-3.20$ & $-3.47$ & fixed\\
TCEDE, $H_0=71$ & $+16.18$ & $+15.66$ & refitted\\
Axion-like EDE best fit & $-8.67$ & $-8.95$ & fixed\\
\colrule
\multicolumn{4}{l}{Dark-atom pipeline (HMcode)}\\
$\lambda=1.050$, synchronous & $-1.27$ & $-1.16$ & line\\
$\lambda=1.077$, synchronous & $-1.67$ & $-1.79$ & refitted\\
$\lambda=1.095$, synchronous & $-1.80$ & $-1.81$ & line\\
$\lambda=1.050$, exact copy & $-2.83$ & $-3.06$ & line\\
$\lambda=1.077$, exact copy & $-3.30$ & $-3.52$ & refitted\\
$\lambda=1.095$, exact copy & $-3.71$ & $-3.96$ & line\\
one field & $+3.13$ & $+3.57$ & line\\
two fields, exact copy & $-0.70$ & $-0.68$ & line\\
\end{tabular}
\end{ruledtabular}
\end{table}

\emph{Precision.} With the perturbation tolerance, multipole spacing and $k$ sampling tightened in two steps, $\DS$ changes by $-0.03$ and $-0.05$ for the synchronous fit at $\lambda=1.095$, by $-0.02$ and $-0.05$ for the exact-copy fit at $\lambda=1.077$ and by $-0.03$ and $-0.03$ for the two-field fit with the modulus in the code. These tests were made at the points of the fits before the restarts described above, which lowered them by at most $0.46$. 

\emph{Range of the lensing potential.} CLASS v3.1 computes the lensing potential on the wavenumber grid of the CMB spectra, which ends at $k_{\max}=2.4\,\ell_{\max}/\tau_0\approx1.6\,{\rm Mpc}^{-1}$; later versions integrate to about $9.5\,{\rm Mpc}^{-1}$. With HMcode the default range gives $4\%$, $10\%$ and $18\%$ less lensing power than the TCEDE pipeline at $L=2000$, $3000$ and $4000$. With the range extended to $9.5\,{\rm Mpc}^{-1}$ the two codes agree to $0.3\%$ in $C_L^{\phi\phi}$ up to $L=8000$ and to $0.01$ in $S$ (Sec.~II\,C). That setting exceeds our memory budget, so the fits were re-evaluated with $k_{\max}\approx5.3\,{\rm Mpc}^{-1}$ (\texttt{k\_scalar\_max\_tau0\_over\_l\_max}~$=8$), which gives $91\%$ of the change in $S$ at that point. At the \LCDM{} best fit of the dark-atom pipeline, at fixed parameters, it changes $S$ by $+1.1$ and the ACT DR6 term by $+1.2$; the ACT lensing likelihood changes by at most $0.03$. It changes $\DS$ by $-0.08$ and $-0.17$ for the synchronous and exact-copy fits at $\lambda=1.077$, and by $-0.09$ and $-0.08$ for the one-field fit with $\Gamma=m/8\pi$ and the two-field fit with the modulus in the code; with the full range these would be about $10\%$ larger.

\section{Cross-check with a proxy dark plasma}\label{app:proxy}
As a check that does not rely on \texttt{class\_adm}, we fitted the symmetry direction with stock
CLASS v3.3.4 and its module for the effective theory of structure formation (ETHOS) for interacting dark matter and dark
radiation~\cite{CyrRacine2016,Archidiacono2019}. The dark baryons are interacting dark matter with
$\omega_{\rm idm}=(\lambda^2-1)\,\omega_b$, and the dark photons are interacting dark radiation with
$N_{\rm idr}=(\lambda^2-1)/0.2271$ and temperature index $n=2$. A patch sets the dark-photon opacity to
$1.01$ times the rescaled visible Thomson opacity at all redshifts, so the dark plasma decouples in
step with visible recombination. There is no dark polarization, and the angular coefficients of the
dark-radiation hierarchy are the ETHOS defaults. The extra free-streaming radiation is
$N_{\rm ur}=2.0308+3.044(\lambda^2-1)$, the massive neutrino ($0.06$\,eV) is not rescaled, the constants
return in a step at $z=50$, and non-linear power uses HMcode. The data are those of
Sec.~II without the supernovae. At each $\lambda$ we took the best of two or three Cobaya
minimizations, without polish; repeated runs scatter by about $0.5$.

The fits are shown in Fig.~1. For $\lambda=1.04$, $1.08$, $1.12$, $1.16$, $1.24$, $1.32$
and $1.45$ they give $H_0=71.3$, $74.1$, $76.9$, $79.9$, $85.7$, $91.4$ and $100.7\kms$, with $\DS=-1.3$,
$-2.0$, $-2.8$, $-3.5$, $-3.3$, $-2.3$ and $+1.3$; a fit at $\lambda=1.001$, a null test of the
minimizer, gives $\DS=-0.4$. Along the whole profile $\Omega_m=0.300\pm0.002$, so DESI is
fitted equally well at every $\lambda$, and $S_8$ rises only from $0.815$ to $0.829$. With a Gaussian
prior $H_0=73.04\pm1.04\kms$ in place of the supernovae, \LCDM{} moves only to $H_0=68.6\kms$ and pays
$18$ in the prior, which the symmetry direction avoids. The proxy and the dark-atom implementation
agree in their totals where they overlap, but they are not independent: they share the likelihood
code, the calibration convention and the CLASS code base.

\section{Varying constants in BBN}\label{app:bbn}
We extended the public code PRyMordial~\cite{Burns2024}, in the version of its
repository\footnote{\url{https://github.com/vallima/PRyMordial}} identified by the git commit \texttt{725d8a8}, as follows.
\begin{enumerate}
\item \emph{Electron mass.} $m_e$ enters the kinematics of the weak rates, whose normalization scales
as $m_e^5$ with the neutron lifetime matched to its present value, and the thermodynamics of the
$e^\pm$ plasma. In the runs with the kick the mass follows the field and the entropy of the plasma is conserved; this changes D/H by $-0.3\%$ and $Y_p$ by $+0.0003$ relative to a plasma with the standard mass.
\item \emph{Fine-structure constant.} $\alpha$ enters the electromagnetic part of the neutron--proton
mass difference, for which we use the lattice value $Q_{\rm QED}=-1.00$\,MeV scaled with $\alpha$; the
Coulomb corrections to the weak rates; charged-particle reactions, through the Gamow factor of the
entrance channel and one further power of $\alpha$ from the normalization of the Coulomb wave function;
and radiative captures, which carry one more power of $\alpha$. The responses are
$\partial\ln(\mathrm{D/H})/\partial\ln\alpha=3.9$ and $\partial\ln Y_p/\partial\ln\alpha=2.5$. The
latter is larger than the $1.9$ of Ref.~\cite{Dent2007} because that work, like Ref.~\cite{Coc2007}, uses
$Q_{\rm QED}=-0.76$\,MeV; with that value we obtain $1.9$.
\item \emph{Coulomb binding.} Coulomb contributions to nuclear binding ($a_C=0.551$\,MeV) enter the
Q-values and nuclear statistical equilibrium. At the final shift of the benchmark they change D/H by $0.02\%$ and $Y_p$ by less than $10^{-4}$; they are included in the runs with constant shifts and omitted in the runs with the kick.
\item \emph{Time dependence.} For the kick, the neutron--proton rates are tabulated at five fractions
of the final shift and interpolated along the field solution, and the nuclear rates are rescaled at
each temperature.
\item \emph{Mirror sector and relic.} A mirror sector present at BBN is evolved as a thermal copy with
its own $e'^\pm$ annihilation. A relic that decays after BBN is included as a matter component that
decays at $t=\tau_X$, or, for the comparison in Sec.~VII\,C, as an exponentially decaying
component together with its decay radiation.
\item \emph{Scans.} In the runs with constant shifts, $\alpha$ and $m_e$ are multiplied by
$1+\delta\alpha/\alpha$ and $1+\delta m_e/m_e$ throughout BBN; the fixed-binding scans impose
$(1+\delta\alpha/\alpha)^2(1+\delta m_e/m_e)=1$ exactly. Each best fit is the minimum of $\chi^2$ along a
scan in steps of $0.25\%$ in $\alpha$ around the minimum ($0.5\%$ for the widest scan), with $Y_p$ and
D/H interpolated by cubic polynomials, whose largest residuals are $3.0\times10^{-6}$ in $Y_p$ and
$5.1\times10^{-5}$ in $10^5\,$D/H.
\end{enumerate}
We neglect finite-radius corrections to the Coulomb penetration, exit-channel barriers, the $\alpha$
dependence of the strong-interaction parts of the binding energies, and resonance energies. All runs cap the
relative tolerance of the PRyMordial integrators at $10^{-7}$; with the default tolerance of its
low-temperature network, D/H carries numerical noise of about $0.2\%$.

\section{Neutrino energy fractions of mirror cascades}\label{app:cascade}
For a relativistic parent we use mean rest-frame energy fractions with Particle Data Group branching
ratios~\cite{PDG2024}. In muon decay the electron carries $\tfrac{7}{20}m_\mu$ on average, so the
neutrino fraction of a muon is $f_\mu=0.65$. This gives $f_{\pi^\pm}=0.725$, $f_{K^\pm}=0.675$,
$f_{K_L}\simeq f_{K_S}\simeq0.50$, $f_\eta=0.12$ and about $0.3$ for nucleon--antinucleon annihilation.
For the $\tau$ we sum over the leptonic modes ($0.65$ and $0.88$), $\pi\nu$ ($0.86$), $\rho\nu$
($0.62$) and the remaining hadronic modes ($\approx0.62$), which gives $f_\tau=0.69$. Light-quark jets
are built from the inclusive energy shares measured in $Z$ decays at the Large Electron--Positron collider (LEP), with resonance decays folded
in: $50\%$ $\pi^\pm$, $25\%$ $\pi^0$, $8\%$ $K^\pm$, $8\%$ $K^0$, $3\%$ $\eta$ and $6\%$ nucleons. Gauge-
and Higgs-boson-like channels are weighted by their Standard Model branching ratios. The dominant
uncertainties are the jet composition and the neglect of energy-dependent decay kinematics, which we
estimate at $\pm0.03$ together. The resulting fractions of the decay energy that end in free-streaming
mirror neutrinos are $F=0.48$ for quark jets, $0.46$ for dark-photon-like decays,
$0.52$--$0.58$ for $W'$-, $Z'$- and Higgs-like decays, $0.69$ for $\tau'$ pairs, $0.65$ for $\mu'$
pairs, and zero for decays into mirror electrons or photons.

These are vacuum values: they assume that every unstable particle decays before it loses energy. A low
mirror baryon density does not guarantee this, because a relativistic charged particle also loses energy
to the mirror photons by inverse Compton scattering. For a muon with Lorentz factor $\gamma$ in a
blackbody of temperature $T'$, in the Thomson limit and with the mirror particle masses equal to the
visible ones, the loss time is $t_{\rm IC}\simeq(1.98\,{\rm s}/\gamma)(T'/1\,{\rm keV})^{-4}$ and
the decay time $2.2\times10^{-6}\,\gamma$\,s. They are equal at $\gamma\simeq38$, an energy of
$4$\,GeV, for $T'=5$\,keV, and at $100$\,GeV for $T'=1$\,keV; for charged pions, which live
less long, the corresponding energy at $5$\,keV is $70$\,GeV. The vacuum fractions therefore apply
when the muons of the cascade are slower than this, which depends on the mass of $X$, on the lifetime and
on how much of the mirror radiation is already present; we have not computed a cascade with losses. In
the opposite limit, in which every muon gives its energy to the mirror photons before it decays, a
charged pion passes $0.21$ of its energy to neutrinos instead of $0.725$, and the fractions become
$F=0.17$ for quark jets and $0.21$--$0.33$ for weak-boson- and Higgs-like decays. Losses thus
lower $F$, and the symmetric value $0.409$ lies between the two limits for the weak-boson-like
channels. The fits at imposed $F$ (Sec.~VII\,B) do not depend on this.

\section{Best-fit parameters}\label{app:bestfit}
Table~\ref{tab:bestfit} lists the parameters of the main fits without the SH0ES calibration. Search ranges
were wide enough that no quoted point lies on a bound.

\begin{table*}[t]
\caption{Best-fit parameters, without the SH0ES calibration. TCEDE pipeline: one $0.05$\,eV neutrino
and linear lensing power, except in the rows marked HMcode (version 2020); the TCEDE best fit has
$\varepsilon=200$ and $\phi_i^2/\Mpbar^2=0.34$ (with HMcode, $210$ and $0.26$), and the fit at $H_0=71\kms$ has $173$ and $0.72$. The axion-like EDE best fit has
$f_{\rm EDE}=0.094$, $\log_{10}z_c=3.55$ and $\theta_i=2.70$, and $0.090$, $3.54$ and $2.71$ with HMcode. Dark-atom pipeline: one $0.06$\,eV neutrino (rescaled as in
Sec.~IV); $\omega_{\rm cdm}$ excludes the dark baryons; synchronous dark constants
unless stated.}
\label{tab:bestfit}
\begin{ruledtabular}
\begin{tabular}{lcccccccc}
Fit & $H_0$ & $\omega_b$ & $\omega_{\rm cdm}$ & $n_s$ & $\ln10^{10}A_s$ & $\tau$ & $A_{\rm act}$ & $P_{\rm act}$\\
\colrule
\multicolumn{9}{l}{TCEDE pipeline}\\
\LCDM & 68.32 & 0.02250 & 0.1179 & 0.9732 & 3.062 & 0.0630 & 1.0029 & 1.0027\\
TCEDE best fit & 69.24 & 0.02264 & 0.1215 & 0.9823 & 3.071 & 0.0621 & 1.0014 & 1.0029\\
TCEDE, $H_0=71$ & 71.00 & 0.02299 & 0.1236 & 0.9987 & 3.087 & 0.0635 & 1.0014 & 1.0032\\
axion-like EDE best fit & 71.00 & 0.02265 & 0.1281 & 0.9846 & 3.071 & 0.0610 & 1.0015 & 1.0025\\
\LCDM, HMcode & 68.36 & 0.02251 & 0.1178 & 0.9741 & 3.058 & 0.0622 & 1.0017 & 1.0025\\
TCEDE best fit, HMcode & 69.17 & 0.02261 & 0.1204 & 0.9809 & 3.064 & 0.0608 & 1.0009 & 1.0028\\
axion-like EDE, HMcode & 70.92 & 0.02265 & 0.1274 & 0.9839 & 3.068 & 0.0604 & 1.0005 & 1.0022\\
\colrule
\multicolumn{9}{l}{Dark-atom pipeline, prescribed constants}\\
\LCDM & 68.26 & 0.02254 & 0.1178 & 0.9736 & 3.057 & 0.0608 & 1.0026 & 1.0026\\
$\lambda=1.050$ & 71.92 & 0.02252 & 0.1302 & 0.9722 & 3.056 & 0.0603 & 1.0023 & 1.0023\\
$\lambda=1.077$ & 73.84 & 0.02250 & 0.1374 & 0.9711 & 3.057 & 0.0604 & 1.0022 & 1.0019\\
$\lambda=1.095$ & 75.16 & 0.02249 & 0.1421 & 0.9705 & 3.057 & 0.0602 & 1.0021 & 1.0020\\
$\lambda=1.077$, exact copy & 73.87 & 0.02257 & 0.1371 & 0.9692 & 3.057 & 0.0595 & 1.0017 & 1.0015\\
\colrule
\multicolumn{9}{l}{Dark-atom pipeline, modulus in the code ($\lambda=1.077$)}\\
one field, $\Gamma=m/8\pi$ & 73.54 & 0.02249 & 0.1374 & 0.9716 & 3.062 & 0.0614 & 1.0034 & 1.0023\\
one field & 73.64 & 0.02249 & 0.1374 & 0.9718 & 3.058 & 0.0601 & 1.0029 & 1.0023\\
two fields & 73.73 & 0.02249 & 0.1372 & 0.9716 & 3.059 & 0.0605 & 1.0030 & 1.0023\\
two fields, exact copy & 73.79 & 0.02259 & 0.1371 & 0.9694 & 3.057 & 0.0594 & 1.0019 & 1.0017\\
two fields, exact copy, $F=0.478$ & 73.77 & 0.02256 & 0.1371 & 0.9699 & 3.059 & 0.0592 & 1.0017 & 1.0015\\
\end{tabular}
\end{ruledtabular}
\end{table*}

\section{Timeline of the scenario and BBN with the kick}\label{app:figs}
\begin{figure*}[t]
\includegraphics[width=\textwidth]{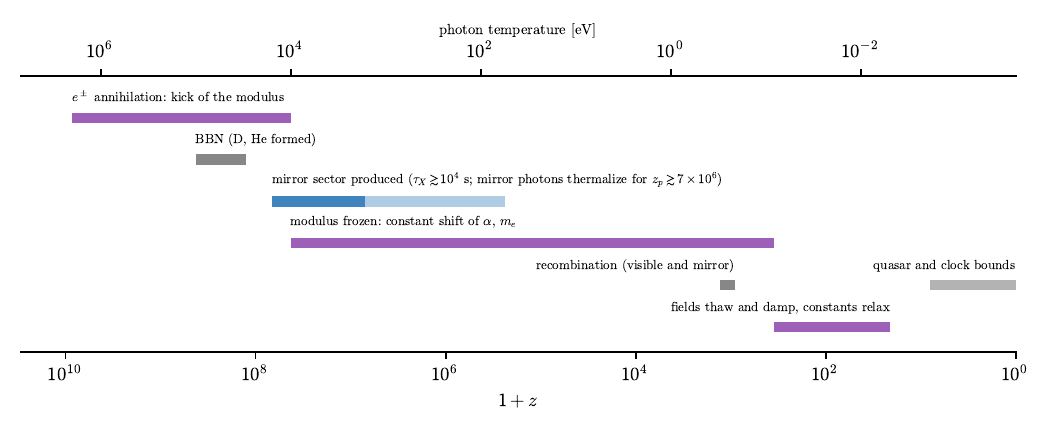}
\caption{Timeline of the scenario: the kick at $e^\pm$ annihilation, BBN, the window for producing the
mirror sector, drawn at the redshift $z_p$ at which the decay rate equals the expansion rate
(Sec.~VII; dark where the mirror photons thermalize, light down to the limit from the
refitted timing test), the constant shift through recombination, the
return of the constants, and the epoch of quasar and atomic-clock bounds.}
\label{fig:timeline}
\end{figure*}

\begin{figure*}[t]
\includegraphics[width=\textwidth]{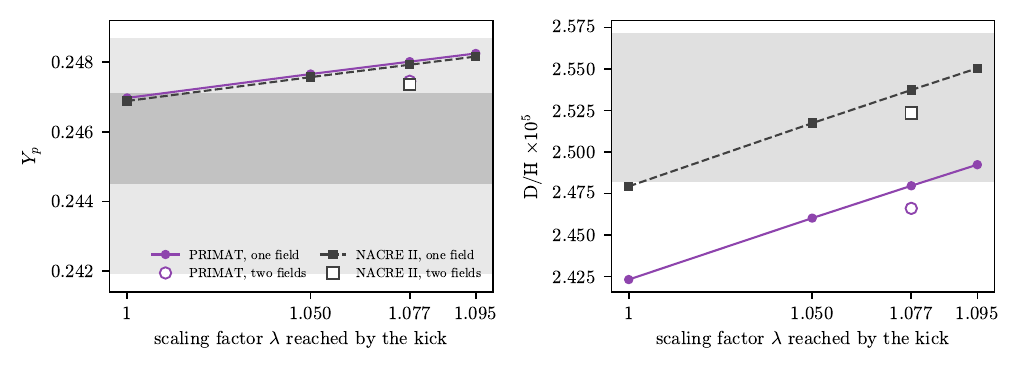}
\caption{Primordial helium and deuterium with the kick and no mirror sector at BBN, for the one-field
couplings that give $\lambda=1$, $1.050$, $1.077$ and $1.095$ and for the two-field benchmark, with
PRIMAT and NACRE~II rates at $\omega_b=0.02250$. Gray bands show the observations; for $Y_p$ the dark
band is the LBT value~\cite{Aver2026} and the light band the earlier one~\cite{Aver2021}; for D/H the
band includes the theory error.}
\label{fig:bbn}
\end{figure*}

Figure~\ref{fig:timeline} places the steps of Part~II in redshift, and Fig.~\ref{fig:bbn} shows the yields of Sec.~VI\,B.

\section{The two-field model: kick and benchmark}\label{app:twofield}
This section gives the details behind the two-field paragraphs of Sec.~VI\,E. The Standard Model couples to the
product $\varphi\sigma$, so that Eq.~(9) holds with $\kappa\varphi/\Mpbar$ replaced by
$\kappa(\varphi/\Mpbar)(\sigma/v_\sigma)$.

The switch field is not a spectator during the kick, because the electron trace sources both fields.
With $s=\sigma/v_\sigma$ and $v=v_\sigma/\Mpbar$,
\begin{align}
u''+\Big(3+\frac{d\ln H}{dx}\Big)u'&=6\kappa\,s\,\frac{\rho_e-3p_e}{\rho_{\rm tot}},\nonumber\\
s''+\Big(3+\frac{d\ln H}{dx}\Big)s'&=\frac{6\kappa}{v^2}\,u\,\frac{\rho_e-3p_e}{\rho_{\rm tot}},
\label{eq:kick2}
\end{align}
and nothing holds $s$ at $1$, because its mass is negligible against $H$ at these temperatures.
Integrating from $u=0$, $s=1$ before annihilation and requiring the same final shift gives
Table~\ref{tab:twofield}. For $v=0.1$ the switch field ends $27\%$ above its initial value,
and the coupling that gives $\lambda=1.077$ is $\kappa=0.093$ instead of $0.106$, so
Eq.~(11) does not hold in the two-field model. For smaller $v$ the two fields
amplify each other. At $v=0.02$ the switch field grows by a factor $3.3$, the required
coupling is half that of the one-field model, and the two fields together store less energy than the
single field does (last column of the table). The shift then scales as $\kappa^{3.9}$, against
$\kappa^2$ for one field. The kick is also later, which moves the BBN yields toward the standard ones. Holding $\sigma$ fixed during annihilation would need a further ingredient that then
releases it; we do not assume one.

\begin{table}[t]
\caption{Kick in the two-field model for $\lambda=1.077$: coupling required, post-kick field values,
fraction of the final shift reached at $T=0.1$\,MeV, and stored energy
$(\varphi^2+\sigma^2)/\varphi_1^2$ for equal masses, where $\varphi_1$ is the post-kick field of the one-field
model (first row).}
\label{tab:twofield}
\begin{ruledtabular}
\begin{tabular}{lccccc}
$v_\sigma/\Mpbar$ & $\kappa$ & $\varphi/\Mpbar$ & $\sigma/v_\sigma$ & at $0.1$\,MeV & energy\\
\colrule
-- & 0.106 & 0.117 & 1 & 49\% & 1\\
0.30 & 0.104 & 0.115 & 1.04 & 47\% & 8.04\\
0.10 & 0.093 & 0.105 & 1.27 & 41\% & 1.98\\
0.05 & 0.077 & 0.092 & 1.74 & 32\% & 1.17\\
0.03 & 0.063 & 0.081 & 2.42 & 26\% & 0.86\\
0.02 & 0.053 & 0.072 & 3.25 & 21\% & 0.69\\
\end{tabular}
\end{ruledtabular}
\end{table}

\emph{A two-field benchmark.} We take $v_\sigma=0.02\,\Mpbar$, equal masses
$m_\sigma=m_\varphi=6\times10^{-30}$\,eV and $\Gamma=0.2\,m$ for both fields, start both fields from
the post-kick values of Table~\ref{tab:twofield}, and evolve them in the Boltzmann code as in
Sec.~VI\,C. These choices keep the energy of the fields small.
\begin{itemize}
\item \emph{History.} The shift of the constants at recombination is the same as in the one-field
model by construction. It relaxes somewhat earlier, because it is proportional to the product of two
fields that thaw together (Fig.~4). The residual variation of $\alpha$ is $6\times10^{-10}$ at
$z=20$ and negligible afterwards.
\item \emph{Energy.} $\varphi$ and $\sigma$ reach $0.32\%$ and $0.26\%$ of the total density
at $z\approx170$, and the decay radiation today has $0.4$ times the CMB photon density.
At the parameters of the Part~I fit, $\theta_s$ shifts by $+0.04\%$, $\sigma_8$ falls by $0.6\%$
and the score rises by $6$. With $v_\sigma=0.05$ and $0.03\,\Mpbar$ at the same masses and
damping the rise is $14$ and $8$, and the shift of the constants scales as
$v_\sigma^{-1.9}$ at fixed coupling, so the benchmark is one point of a family in which the
switch scale, the coupling and the energy of the fields are tied together. Other choices cost more: with $v_\sigma=0.1\,\Mpbar$,
$m_\sigma=10^{-30}$\,eV and $\Gamma=m/8\pi$, the switch field thaws late and decays slowly, and the score
rises by $91$.
\item \emph{Equivalence principle.} Today $\sigma/v_\sigma$ and $\varphi/\Mpbar$ are below $10^{-20}$ (the
integration returns $6\times10^{-23}$ and $1\times10^{-24}$, which is its numerical floor; the damping alone gives far
less), so the linear couplings of both fields to matter are negligible. This
assumes that near the Earth the fields take their cosmological values. Matter does not displace them at
linear order, because the coupling is bilinear; it mixes the two fields, with a length scale of about
$10^{12}$\,m at terrestrial density, far larger than the Earth, so we estimate that the configuration
$\varphi=\sigma=0$ is not destabilized.
\item \emph{BBN.} The kick is later than in the one-field model ($21\%$ complete at $0.1$\,MeV,
against $49\%$). It gives $Y_p=0.2474$ and $\mathrm{D/H}=2.466\times10^{-5}$ with PRIMAT
($\chi^2=3.4$).
\item \emph{CMB.} Refitting gives $\DS=+1.4$ without SH0ES and $-28.3$ with it, at $H_0=73.7\kms$ (Table~IX); with exact-copy dark constants the values are $-0.7$ and $-30.4$, and $-0.5$ and $-30.2$ for the quark-jet composition $F=0.478$.
\end{itemize}
The damping matters for the equivalence principle. If $\sigma$ were undamped, $\varphi$ would stay coupled at $d_e\approx5.5\times10^{-5}$, giving $\eta\approx-8.48\times10^{-15}$; if $\varphi$ were undamped, $\eta\approx-1.04\times10^{-14}$. These lie $2.6\sigma$ and $3.3\sigma$ from the measured central value, so both must be damped for the benchmark to lie within $2\sigma$.

\section{Composition, thermalization and the relic during BBN}\label{app:therm}
This section gives the details behind Secs.~VII\,B and~VII\,C.

\emph{Composition path and decay channels.} At $F=0.478$ the mirror temperature is $3.1\%$ lower and the mirror
baryon density $12\%$ lower than at
$F=0.409$. The cost of $+0.5$ is therefore the cost of the whole path and must not be added to the
temperature tolerance of Sec.~VII\,B, which was measured at fixed composition.

In a generic model this division of the decay energy would be a separate tuning. It is not if $X$
belongs to the mirror sector and decays into a mirror copy of the Standard Model. If every unstable
product decays before it loses energy, the cascade runs as in vacuum, and the share of the energy that
ends in mirror neutrinos is fixed by Standard Model particle physics and the primary channel
(Sec.~\ref{app:cascade}). The fractions that follow are such vacuum values. They hold for pions below about $70$\,GeV, but
for muons only below a Lorentz factor of about $40$ in a mirror photon bath of $5$\,keV, because
faster muons lose their energy to the bath by inverse Compton scattering before they decay; the
fraction then falls below the vacuum value (Sec.~\ref{app:cascade}). Hadronic channels give $F\approx0.48$,
because a charged pion puts about $72\%$ of its energy into neutrinos and a neutral pion none; the refit
at $F=0.478$ shows that this costs only $+0.5$. Weak-boson-like channels lie near $F=0.55$, which
costs $+1.4$. Decays into mirror electrons or photons alone give no free-streaming component and cost
$+316$ at fixed cosmology; they were not refitted, and neither were $\tau'$ and $\mu'$ pairs, with $F=0.69$ and
$0.65$.

\emph{Does the mirror plasma thermalize?} The cascade deposits its electromagnetic energy as energetic
mirror photons and pairs. Compton scattering brings them into kinetic equilibrium, but a blackbody also
requires processes that change the photon number, chiefly double Compton scattering, which in the
visible sector is efficient above $z\simeq2\times10^6$~\cite{Hu1993prd,Chluba2012}. The mirror plasma is
more dilute and colder. For an exact-copy mirror plasma of hydrogen the electron density is
$0.18$ of the visible one and the temperature $\xi$ times the visible one; with the shifted constants
and the faster expansion, the rate of the small-distortion estimate~\cite{Hu1993prd,Chluba2012}, which
scales as $n_e\sigma_T\alpha^{1/2}(T/m_e)^{3/2}/H$, is $0.10$ of the visible one at a given
redshift. A blackbody is then reached only if the energy is released at $z\gtrsim5\times10^{6}$, when
the Universe is younger than about $10^{6}$\,s; for a decaying relic, which releases most of its
energy at an age close to its lifetime, this means $\tau_X\lesssim10^{6}$\,s, or
$z_p\gtrsim7\times10^{6}$. This estimate was derived for a small distortion of an existing blackbody. Here
the mirror sector starts empty, and the cascade must build the photon number as well; while the mean
energy per particle is a sizable fraction of $m_e$, pairs make the number-changing processes fast, but
we have not followed that evolution, so the bound is indicative. For later production the mirror photons
keep a deficit in number: a Bose--Einstein spectrum, hotter than a blackbody of the same energy
density. Their energy density, which is what the symmetry constrains, is unaffected, but mirror
recombination is then delayed, possibly to well after visible recombination. Our fits assume thermal
mirror photons. They show that moving mirror recombination from $z\approx1100$ to
$z\approx1780$ changes the score by only $1.6$, but we have not computed the case of
late mirror recombination. The mirror neutrinos from the cascade are not thermal either. This matters
little: apart from the rescaled massive species of Sec.~IV, the extra free-streaming
radiation is massless in the fits, and for massless particles only the energy density enters the linear
equations.

\emph{The relic during BBN.} Before it decays the relic is matter, and it is not negligible during
nucleosynthesis: for $\tau_X=10^4$\,s it carries $2.6\%$ of the radiation density when
deuterium forms. Added to the one-field kick of Sec.~VI\,B, relics with $\tau_X=10^4$, $10^5$ and $10^6$\,s change
$Y_p$ by $+0.0007$, $+0.0002$ and $+0.0001$ and D/H by $+3.5\%$,
$+1.1\%$ and $+0.4\%$, and give $\chi^2=5.8$, $3.7$ and
$3.8$ with PRIMAT rates, against $4.0$ without the relic. If we ask that the fit be no worse than standard BBN at the same
baryon density ($\chi^2=6.1$), BBN allows
lifetimes down to about $10^4$\,s, where helium is $2.3\sigma$ high and the relic matters as
much as the kick. These runs let the relic decay all at once at $t=\tau_X$. With an exponential decay and its radiation followed, the relic that gives the same final mirror radiation is $13\%$ more abundant; for $\tau_X=10^4$\,s it carries $2.9\%$ of the radiation density when deuterium forms, D/H changes by $+3.8\%$ and $\chi^2=6.4$, just above standard BBN, which moves the lower edge slightly above $10^4$\,s; the mirror radiation already released by then is $0.05\%$ of the radiation density.

\section{TCEDE perturbation equations and numerical tests}\label{app:tcede}

\emph{Model and background.} One neutrino mass eigenstate (bare mass $m_{\nu0}=0.05$\,eV, $g=2$) has
$m_\nu=m_{\nu0}\,[1+\varepsilon\bar\phi^2]$ with $\bar\phi\equiv\phi/\Mpbar$ and
$\bar\beta\equiv d\ln m_\nu/d\bar\phi=\Mpbar\beta=2\varepsilon\bar\phi/(1+\varepsilon\bar\phi^2)$. The other $2.046$
effective neutrino species are massless. Since a homogeneous field leaves the comoving momentum $q$
unchanged, the relic keeps its Fermi--Dirac form $f_0(q)=[e^{q/T_{\nu0}}+1]^{-1}$, and
\begin{gather}
\rho_\nu=\frac{g}{2\pi^2a^4}\!\int\! dq\,q^2E\,f_0,\qquad
p_\nu=\frac{g}{6\pi^2a^4}\!\int\! dq\,\frac{q^4}{E}f_0,\nonumber\\
\mathfrak p_\nu=\frac{g}{6\pi^2a^4}\!\int\! dq\,\frac{q^6}{E^3}f_0,
\end{gather}
with $E=\sqrt{q^2+a^2m_\nu^2(\bar\phi)}$ and $\mathfrak p_\nu$ the pseudo-pressure. In conformal time, with $\mathcal H=a'/a$ and
$T_\nu=\rho_\nu-3p_\nu$,
\begin{align}
\bar\phi''+2\mathcal H\bar\phi'&=-\frac{a^2}{\Mpbar^2}\,\bar\beta\,T_\nu,\nonumber\\
\rho_\nu'+3\mathcal H(\rho_\nu+p_\nu)&=\bar\beta\bar\phi'\,T_\nu,
\end{align}
and the scalar, with $\rho_\phi=\Mpbar^2\bar\phi'^2/2a^2$, carries the opposite exchange term. The
neutrino energy already contains the effective potential, so no thermal potential is added to the
Friedmann equation. The field starts at rest at $z=4\times10^9$, as in Ref.~\cite{Kamionkowski2025}.

\emph{Perturbations.} We work in synchronous gauge with the conventions of Ma and
Bertschinger~\cite{Ma1995}, scalar modes only. Writing $f=f_0(q)[1+\Psi(k,q,\hat n,\tau)]$, expanding
$\Psi$ in Legendre moments $\Psi_\ell$ and defining $D\equiv d\ln f_0/d\ln q$, the coupled species obeys
\begin{align}
\Psi_0'&=-\frac{qk}{E}\Psi_1+\frac{h'}{6}D,\nonumber\\
\Psi_1'&=\frac{qk}{3E}(\Psi_0-2\Psi_2)-\frac{a^2m_\nu^2k}{3qE}\,\bar\beta\,\delta\bar\phi\,D,\nonumber\\
\Psi_2'&=\frac{qk}{5E}(2\Psi_1-3\Psi_3)-\Big(\frac{h'}{15}+\frac{2\eta'}{5}\Big)D,\nonumber\\
\Psi_\ell'&=\frac{qk}{(2\ell+1)E}\big[\ell\Psi_{\ell-1}-(\ell+1)\Psi_{\ell+1}\big]\quad(\ell\ge3).
\end{align}
The scalar force enters only the dipole; there is no extra monopole term, because $q$ is conserved. The
stresses include the explicit mass response,
\begin{align}
\delta\rho_\nu&=\delta\rho_\Psi+\bar\beta\,\delta\bar\phi\,T_\nu,\nonumber\\
\delta p_\nu&=\delta p_\Psi-\bar\beta\,\delta\bar\phi\,(p_\nu-\mathfrak p_\nu),
\end{align}
where $\delta\rho_\Psi$ and $\delta p_\Psi$ are the moments above with $f_0\to f_0\Psi_0$. The momentum
density and shear take their usual forms, with the field-dependent $E$ in every integral. The scalar
obeys
\begin{multline}
\delta\bar\phi''+2\mathcal H\delta\bar\phi'+k^2\delta\bar\phi+\tfrac12h'\bar\phi'\\
=-\frac{a^2}{\Mpbar^2}\Big[\bar\beta\,\delta T_\nu+\frac{d\bar\beta}{d\bar\phi}\,T_\nu\,\delta\bar\phi\Big],
\end{multline}
with $\delta T_\nu=\delta\rho_\nu-3\delta p_\nu$ including the explicit mass terms. Its stresses are
$\delta\rho_\phi=\delta p_\phi=\Mpbar^2\bar\phi'\delta\bar\phi'/a^2$ and
$(\rho_\phi+p_\phi)\theta_\phi=k^2\Mpbar^2\bar\phi'\delta\bar\phi/a^2$, with no shear.

\emph{Initial conditions and truncation.} Each mode starts in the adiabatic growing mode deep in
radiation domination, with the coupled neutrino relativistic and
$\delta\bar\phi=\delta\bar\phi'=0$; the driven solution is of higher order in $k\tau$. The coupled
species uses the exact momentum hierarchy at all times, with no fluid approximation, $48$
Gauss--Laguerre momentum nodes and $\ell_{\max}=50$, closed by the standard free-streaming truncation.
Fits use $24$ nodes and $\ell_{\max}=25$. Every quoted score is re-evaluated with $48$ nodes and $\ell_{\max}=50$, twice as many
background steps and a perturbation tolerance five times tighter.

\emph{Tests.} (i)~\emph{Conservation.} At the benchmark of Ref.~\cite{Kamionkowski2025}, whose transient is about
seven times that of our best fit, and with $192$ momentum nodes and $\ell_{\max}=100$, the neutrino and scalar
energy--momentum equations were integrated independently from the printed histories of the code ($k=0.003$--$0.3\,{\rm Mpc}^{-1}$,
$1080\le z\le10^7$). The 95th-percentile relative residuals are $8.3\times10^{-8}$ in the neutrino energy
equation, $1.4\times10^{-7}$ in the neutrino momentum equation, $3.5\times10^{-8}$ in the total energy equation, $5.9\times10^{-8}$ in the
Einstein pressure equation and $1.8\times10^{-5}$ in the total momentum equation, which arises at the largest scale tested, $k=0.003\,{\rm Mpc}^{-1}$, and at $1080<z<5000$, does not change when the neutrino momentum nodes, multipoles and integration tolerance are raised from $(32,17,10^{-7})$ to $(192,100,2\times10^{-8})$, and is the same with the coupling switched off. An independent background solver agrees with the code to
$2.1\times10^{-7}$ in $H(z)$.
(ii)~\emph{The uncoupled limits.} With $\bar\phi_i=0$ at the \LCDM{} best-fit cosmology the likelihood
pipeline returns the \LCDM{} score exactly, with identical lensed spectra. With $\varepsilon=0$ and
$\bar\phi_i\neq0$ the transfer functions are identical to those of the uncoupled code.
(iii)~\emph{Momentum sampling, hierarchy and precision.} Going from the settings of the fits to those of the re-evaluation
changes $S$ by at most $0.04$ at every quoted \LCDM{} and TCEDE point, and $\DS$ of the best fit by $0.02$. Starting every Fourier mode ten times earlier than the CLASS defaults changes $S$ at the best fit by $7\times10^{-5}$.
(iv)~\emph{Coupled perturbations where lensing is sourced.} The axion-like control tests the likelihood
pipeline but not the coupled perturbations, so we checked their size directly. At the redshifts and
scales that source most of the CMB lensing the coupling is weak: at the best fit the scalar force between
coupled neutrinos is at most $0.7$ times gravity for $z\le5$ and acts on $0.4\%$ of the matter. The
density contrast of the coupled neutrino then differs from that of an uncoupled $0.05$\,eV neutrino at
the same parameters by less than $0.7\%$ ($1.4\%$ at $H_0=71\kms$) for
$k=0.003$--$1\,h\,{\rm Mpc}^{-1}$, and the neutrino supplies at most $0.3\%$ of the density
perturbation. Between $z=5$ and $100$, which supplies $7\%$ of the lensing kernel at $L=30$ and
$35\%$ at $L=1000$, we sampled $z=10$, $20$ and $50$: the force reaches $5$ times gravity
($9$ at $H_0=71\kms$), the contrast differs by at most $3\%$ ($6\%$), and the
neutrino still supplies at most $0.3\%$ of the density perturbation. The coupling is strong only
earlier: at the redshifts
that we sampled in $100\le z\le1100$ the scalar force is $10$ to $390$ times gravity. There the contrast departs from that
of an uncoupled neutrino by $3$--$14\%$ at $z=100$--$800$ and by $35\%$ at $z=1100$
($6$--$21\%$ and $70\%$ at $H_0=71\kms$), where the neutrino is still
$75\%$ heavier than today and supplies up to $5\%$ of the matter density perturbation;
these redshifts contribute less than $0.4\%$ of $C_L^{\phi\phi}$. Weighted by the lensing kernel over all the
sampled redshifts, of which $z\le5$ supplies $93\%$ at $L=30$ and $64\%$ at $L=1000$, the coupled
terms change $C_L^{\phi\phi}$ by less than $4\times10^{-5}$ ($9\times10^{-5}$ at $H_0=71\kms$) for $L\le1000$; in
each interval of redshift the bound takes the larger of the two end values of the change of the source. The
linear rms of the neutrino contrast
stays below $0.05$ at all redshifts. The lensing scores therefore rest on the background transient and
on standard clustering; the coupled perturbations matter for the primary anisotropies, which tests
(i)--(iii) cover.

\bibliography{refs}

\end{document}